\documentclass[12pt]{article}
\usepackage{amsmath}
\usepackage{graphicx,psfrag,epsf}
\usepackage{enumerate}
\usepackage{natbib}
\usepackage{url} 

\newcommand{\blind}{1}

\usepackage[utf8]{inputenc} 

\usepackage{geometry} 
\usepackage[]{rotating}
\usepackage{amssymb}

\usepackage{booktabs} 
\usepackage{array} 
\usepackage{paralist} 
\usepackage{verbatim} 
\usepackage{subfig} 
\usepackage{float}
\usepackage{natbib}

\usepackage{sectsty}
\allsectionsfont{\sffamily\mdseries\upshape} 

\usepackage[titles,subfigure]{tocloft} 

\newcommand{\openr}{\hbox{${\rm I\kern-.2em R}$}}
\newcommand{\openn}{\hbox{${\rm I\kern-.2em N}$}}

\usepackage{graphicx}
\usepackage[font=singlespacing]{caption}
\newcommand\independent{\protect\mathpalette{\protect\independenT}{\perp}}
\def\independenT#1#2{\mathrel{\rlap{$#1#2$}\mkern2mu{#1#2}}}
\usepackage{mathrsfs}
\usepackage{tabulary}
\usepackage{enumerate}
\usepackage{setspace}
\usepackage{multirow}
\usepackage{listings}
\usepackage[T1]{fontenc}
\usepackage[usenames,dvipsnames]{color}

\usepackage{authblk}

\title{Complier stochastic direct effects: identification and robust estimation}

\date{}
\begin{document}


\def\spacingset#1{\renewcommand{\baselinestretch}%
{#1}\small\normalsize} \spacingset{1}


\if1\blind
{
  \title{\bf Complier stochastic direct effects: identification and robust estimation}
  \author{\small{Kara E. Rudolph\thanks{
    The authors gratefully acknowledge \textit{R00DA042127} (PI: Rudolph)}\hspace{.2cm}\\
    Department of Emergency Medicine, University of California, Davis and Division of Epidemiology, University of California, Berkeley\\
    and \\
    Oleg Sofrygin\\
    Division of Biostatistics, University of California, Berkeley\\
    and \\
    Mark J. van der Laan\\
    Division of Biostatistics, University of California, Berkeley}}
  \maketitle
} \fi

\if0\blind
{
  \bigskip
  \bigskip
  \bigskip
  \begin{center}
    {\LARGE\bf Complier stochastic direct effects: identification and robust estimation}
\end{center}
  \medskip
} \fi

\begin{abstract}
Mediation analysis is critical to understanding the mechanisms underlying exposure-outcome relationships. In this paper, we identify the instrumental variable (IV)-direct effect of the exposure on the outcome not through the mediator, using randomization of the instrument. To our knowledge, such an estimand has not previously been considered or estimated. We propose and evaluate several estimators for this estimand: a ratio of inverse-probability of treatment-weighted estimators (IPTW), a ratio of estimating equation estimators (EE), a ratio of targeted minimum loss-based estimators (TMLE), and a TMLE that targets the CSDE directly. These estimators are applicable for a variety of study designs, including randomized encouragement trials, like the MTO housing voucher experiment we consider as an illustrative example, treatment discontinuities, and Mendelian randomization. We found the IPTW estimator to be the most sensitive to finite sample bias, resulting in bias of over 40\% even when all models were correctly specified in a sample size of N=100. In contrast, the EE estimator and compatible TMLE estimator were far less sensitive to finite samples. The EE and TMLE estimators also have advantages over the IPTW estimator in terms of efficiency and reduced reliance on correct parametric model specification.
\end{abstract}

\noindent%

\spacingset{1.45} 


\section{Introduction}
Mediation analysis is critical to understanding the mechanisms underlying exposure-outcome relationships. It can be used to decompose the total effect into its path-specific effects---usually categorized as direct effects, meaning the effect of the exposure on the outcome not operating through a mediator, and indirect effects, meaning the path from the exposure to the mediator to the outcome \citep{ogburn2012commentary}. For example, such decomposition has led to understanding of which locations in the brain are responsible for transmitting pain \citep{chen2015high} and mechanisms underlying associations between early life body size and breast cancer \citep{rice2016mammographic}. Such scenarios reflect observed data $O=(W,Z,M,Y)$, where $W$ are covariates, $Z$ is exposure, $M$ is mediator, and $Y$ is outcome. 

Less research has been devoted to estimating path-specific effects where there is an instrument for the exposure, reflecting observed data $O=(W,A,Z,M,Y)$, where $A$ is an instrument for the overall effect of $Z$ by only affecting $M$ and $Y$ through $Z$ and satisfies the econometric criteria for an instrument \citep{joffe2008extended}. In this paper, we consider such a data structure and are concerned with estimating the path-specific direct effect of $Z$ on $Y$, not through $M$, using instrument $A$ to address observed and unobserved confounding of the exposure-outcome relationship. Such an estimand would be an instrumental variable (IV)-direct effect or complier direct effect. To our knowledge, such an estimand has not previously been considered or estimated, though we review related work below. 

Recent work considering the same observed data structure $O=(W,A,Z,M,Y)$ has identified and estimated stochastic direct and indirect effects of the \textit{instrument} on the outcome not operating through a mediator in the direct effects case and operating through a mediator in the indirect effects case \citep{rudolph2017robust}, treating $Z$ as a time-varying confounder \citep{didelez2006direct,vanderweele2016mediation,zheng2017longitudinal}. Joffe et al. also considered observed data $O=(W,A,Z,M,Y)$ but where $Z$ and $M$ are sequential exposures, with $A$ being an instrument for each \citep{joffe2008extended}. In this case, $A$ affects $Z$ and $M$ but not $Y$. The authors were concerned with estimating the overall effect of $Z$, because it could no longer be identified using standard IV approaches. More recently, Frolich and Huber considered observed data $O=(W,A_1,A_2,Z,M,Y)$ with $A_1$ and $A_2$ being distinct instruments for $Z$ and for $M$, respectively \citep{frolich2017direct}. These authors demonstrated how one can identify IV mediation estimands for the direct effect of $Z$ on $Y$ not through $M$ and for the indirect effect of $Z$ on $Y$ through $M$ using these two distinct instruments. Other work has considered another instrumental variable observed data structure, $O=(W,Z,M,Y)$, where $Z$ and $W$ interact together to form an instrument for $M$, relaxing the sequential ignorability assumption \citep{ten2007causal,dunn2007modelling,albert2008mediation,small2011mediation}. However, again, to our knowledge, there has been no research on the identification or estimation of IV mediation estimands in the observed data structure $O=(W,A,Z,M,Y)$, where $A$ is an instrument for the total effect of exposure $Z$, which in turn may affect mediator $M$ and outcome $Y$, and where $A$ adheres to the exclusion restriction assumption of instruments, so does not directly affect either $M$ or $Y$. 

We address this research gap by identifying an IV causal quantity of the direct effect of the exposure on the outcome (the effect of $Z$ on $Y$ not through $M$) (the unmediated portion of the effect), using randomization of the instrument, $A$. We call this estimand the complier stochastic direct effect (CSDE). We propose and evaluate several estimators for this estimand: 1) an inverse-probability-of-treatment weighted estimator (IPTW), 2) an 
 estimating equation estimator (EE), and 3) a targeted minimum loss-based estimator (TMLE). 
 Both the EE and TMLE estimators are robust to several combinations of model misspecifications, the details of which are described in a later section. 
 In contrast, the IPTW estimator may not be consistent if the instrument or mediation models are incorrectly specified. 

The paper is organized as follows. In Section 2, we introduce notation and the structural causal model representing our data structure. In Section 3, we define the causal quantity of interest, the CSDE, and establish its identification from the data distribution under specified assumptions. Sections 4.1, 4.2, and 4.3 detail the IPTW, EE, and TMLE estimators, respectively. Each of the estimator sections is written to stand alone, so one can read whichever section is of interest without compromising understanding. Section 5 presents the simulation study that demonstrates each estimator's consistency, efficiency, and robustness properties in finite samples. In Section 6, we apply these estimators to a real-world instrumental variable scenario where we estimate the direct effect of using a Section 8 housing voucher to move out of public housing on subsequent adolescent substance use outcomes not mediated by parental mental health, employment, and parent-child closeness using randomization of housing voucher receipt as the instrument. Section 7 concludes.\\


\section{Notation and Structural Causal Model}
\label{notationsection}
We observe data $O=(W,A,Z,M,Y)$ for each of $n$ individuals, where we assume $O_1, ..., O_n$ are i.i.d. for the true, unknown data distribution, $P_0$ on $O$. The subscript 0 denotes values under this true, unknown distribution $P_0$. $P$ is any probability distribution in statistical model, $\cal{M}$, which is the set of distributions for which our estimand is identifiable and is discussed further below and in the following Section. Values are a particular $P$ are not given subscripts. The subscript $n$ denotes estimates. 

$W$ is a vector of exogenous baseline covariates, $W=f(U_W)$, where $U_W$ is unobserved exogenous error on $W$ \citep{pearl2009causality}. We consider the statistical model, $\cal{M}$, where $A$ is a binary instrument (with the attendant assumptions of instrumental variables \citep{angrist1996identification}) of binary exposure $Z$, with 
 $A=f(W,U_A)$
 and $Z=f(W, A, U_Z)$, where again, $U_A$ and $U_Z$ are unobserved exogenous errors. $M$ is a binary mediator, with $M=f(W, Z, U_M)$, and $Y$ is an outcome, $f(W,Z,M,U_Y)$, where $U_M$ and $U_Y$ represent exogenous errors. Adhering to the constraints of our statistical model in which $A$ is an instrument, $Y$ does not depend on $A$ conditional on $Z$, and $M$ does not depend on $A$ conditional on $Z$. This is equivalent to exclusion restriction assumption of instruments \citep{angrist1996identification}.
 However, the estimand and estimation approaches we consider also work in the scenario where $M$ may depend on $A$ conditional on $Z$: $M=f(W, A, Z, U_M)$. We describe differences in the estimator details for such a scenario in the Web appendix. In this alternative scenario, $A$ is not an instrument for the total effect of $Z$ on $Y$, and the estimation approach suggested by \cite{joffe2008extended} would also be appropriate.
 
The density of the true distribution $P_0$ of $O$, $p_0(O)$ can be factorized as $$ p_0(O) = p_0(Y | W,Z,M)p_0(M|W,Z)p_0(Z|W,A)p_0(A|W)p_0(W).$$
We note that an identification assumption of monotonicity of $A$ on $Z$, detailed in the next section, places an additional constraint on the statistical model.

\section{Complier Stochastic Direct Effect Estimand and Identification} 

Our causal quantity of interest is the CSDE, which we define as 
 $\psi_{CSDE}=E(Y_{Z=1, g^*_{M|0,W}} - Y_{Z=0, g^*_{M|0,W}} | Z_1 -Z_0 =1)$, where for each $a \in \{0,1\}$, $Z_a$ indicates the counterfactual exposure that would be observed if instrument $A=a$ were assigned and where for each $z \in \{0,1\}$, $Y_{z, g^*_{M|0,W}}$ indicates the counterfactual outcome that would be observed if exposure $Z=z$ were assigned and under a given stochastic intervention on the mediator $g^*_{M|0,W}$, where the user sets $M$ equal to $m$ with probability $P(M=m | A=0, W=w)$. We note that this stochastic intervention marginalizes over $Z$, and also note that $g^*_{M|0,W}$ can be set equal to the true distribution, $g_{M|0,W, \text{ } 0}$, or a data-dependent version estimated from the observed data, $\hat{g}_{M|0,W}$.

The statistical parameter, $\Psi_{CSDE}$, is a mapping $\Psi_{CSDE}:\cal{M} \rightarrow R$ that maps a probability distribution $P$ in our statistical model $\cal{M}$ to a real number $R$. $\Psi_{CSDE}(P) = \frac{\Psi_{SDE}(P)}{\Psi_{FS}(P)}$, 
 where $\Psi_{SDE}$ is the statistical parameter for the stochastic direct effect (SDE) of $A$ on $Y$ given by \begin{equation}
     \begin{aligned}
    \Psi_{SDE} &\equiv E_0(E_0(E_{g^*_{M|0, W}}\{E_0(Y | W,Z, M) | W,Z \}| W,A=1) )\\
     & - E_0(E_0(E_{g^*_{M|0, W}}\{E_0(Y | W,Z, M) | W,Z \}| W,A=0)),
    \end{aligned}
 \end{equation} and where $\Psi_{FS}$ is the statistical parameter for the first-stage (FS) effect of $A$ on $Z$ given by  \begin{equation} \Psi_{FS} \equiv E_0(E_0(Z | W, A=1)) - E_0(E_0(Z | W, A=0)). \end{equation}
The causal quantity $\psi_{CSDE}$ is identified by the statistical parameter $\Psi_{CSDE}$, \begin{equation}
    \begin{aligned}
   \psi_{CSDE}= E(Y_{Z=1, g^*_{M|0,W}} - Y_{Z=0, g^*_{M|0,W}} | Z_1 -Z_0 =1)  \equiv \Psi_{CSDE} &= \Psi_{SDE}/\Psi_{FS},
    \end{aligned}
\end{equation} under the assumptions enumerated below. The proof for the identification result is in the Web appendix. 

The assumptions needed for identifiability are: \begin{enumerate}
\item Sequential randomization: $A \perp Z_a |W$, $A \perp Y_{a,g^*_{M|0,W}}|W$, and $M \perp Y_{a,g^*_{M|0,W}}|W$ 
\item $Y_{a,z} = Y_z,$ which is the exclusion restriction assumption, stating that the instrument $A$ only affects the outcome $Y$ through the exposure $Z$,
\item $Z_1 - Z_0 \ge 0,$ which is the monotonicity assumption, meaning that the instrument $A$ cannot decrease exposure, 
\item Positivity assumptions: $P_0(A=a\mid W)>0$ for all $a \in A$, 
 and $\frac{g^*_{M | 0,W}(m|W)}{P(M=m|Z,W)}< \infty$ a.e. which can also be written, 
$P(M = m|Z,W) > 0$ for all $m$ in the support of $g^*_{M|0,W}(m|W)$, i.e., all $m$ s.t., $g^*_{M|0,W}(m|W)>0$, 
 and 
\item $E_0(Z_1 - Z_0 | W) \ne 0$, which means that the average effect of the instrument on the exposure does not equal 0. 
\end{enumerate}

\section{Estimators}
We now describe several estimators of a data-dependent version of the CSDE parameter that assumes a known stochastic intervention on $M$ estimated from the observed data, which we denote $\hat{g}_{M|0,W}$. So, here $g^*_{M|0,W}$=$\hat{g}_{M|0,W}$. 
 In the first subsection, we describe several estimators that estimate $\Psi_{CSDE}$ by estimating the numerator and denominator separately: a ratio of IPTW estimators, a ratio of EE estimators, and two ratios of TMLE estimators. In the second subsection, we describe a TMLE that targets the CSDE ratio itself, thus making a compatible plug-in estimator.
 
\subsection{Estimators that estimate the numerator and denominator separately}
\subsubsection{Inverse Probability of Treatment Weighted Estimator}

We first describe how to compute $\Psi_{CSDE}$ by using an IPTW estimator of the numerator, $\Psi_{SDE}$, and denominator, $\Psi_{FS}$, separately. 
 The R code to program this estimator is included in the supplementary Web appendix. 

The inverse probability of treatment weights for estimating $\Psi_{SDE}$ 
 are \begin{equation} IPTW_{SDE}=\frac{(2A-1)\hat{g}_{M|0,W}}{g_{A|W}g_{M|Z,W} }. \end{equation}
 Let $g_{A,n}$ and $g_{M,n}$ be estimators of $g_{A|W}=P(A=a | W)$ and $g_{M|Z,W}=P(M=m | Z,W)$, respectively. $g_{A,n}$ can be estimated by predicted probabilities from a logistic regression model of $A$ on $W$. One could use machine learning in model fitting but we will describe estimation in terms of parametric model fitting for simplicity. $g_{M,n}$ can be estimated by 
  predicted probabilities from a logistic regression model of $M$ on $W, Z$. $\hat{g}_{M|0,W}$ is treated as known, estimated from the observed data, marginalizing out $Z: \sum\limits^1\limits_{z=0} P(M=m|Z=z,W)P(Z=z|A=0,W)$ \citep{vanderweele2016mediation}. The IPTW estimate of $\Psi_{SDE}$ is the empirical mean of outcome, $Y$, weighted by an estimate of $IPTW_{SDE}$.

The inverse probability of treatment weights for estimating $\Psi_{FS}$ 
 are $IPTW_{FS}=\frac{2A-1}{g_{A|W}}$, where $g_{A,n}$ is estimated as above. The IPTW estimate of $\Psi_{FS}$ is the empirical mean of $Z$, weighted by an estimate of $IPTW_{FS}.$

The ratio of these two IPTW estimates 
 gives the IPTW estimate of parameter $\Psi_{CSDE}$. The associated variance can be estimated as the sample variance of the estimator's influence curve (IC), which is 
\begin{equation}
D_{IPTW}(P) = \frac{D_{IPTW_{SDE}}(P)}{\Psi_{FS}(P)} - \frac{\Psi_{SDE}(P)D_{IPTW_{FS}}(P)}{\Psi_{FS}^2(P)}, \end{equation}
and where 
\begin{equation}
D_{IPTW_{SDE}}(P) = \frac{(2A-1)\hat{g}_{M|a^*,W}}{g_{A|W}g_{M|Z,W}} Y - \Psi_{SDE}
\end{equation}
and where
\begin{equation}
D_{IPTW_{FS}}(P) = \frac{2A-1}{g_{A|W}}Z - \Psi_{FS}.
\end{equation}

We note that the above is the influence curve using true $g_{A|W}$ and true $g_{M|Z,W}$. If we use parametric models and maximum likelihood estimates of $g_{A|W}$ and $g_{M|Z,W},$ then the sample variance of the above influence curve will be conservative.

\subsubsection{Estimating Equation Estimator}

We now describe how to estimate the $\Psi_{CSDE}$ by using an EE estimator of the numerator, $\Psi_{SDE}$, and denominator, $\Psi_{FS}$, separately. The efficient influence curve we detail for $\Psi_{SDE}$ is novel in that it respects the constraints on our statistical model---namely the exclusion restriction and monotonicity assumptions necessary for identification. These EE estimators use the same estimator of the conditional distribution of $Z$ given $A$ and $W$ for both the numerator and denominator. 
 The R code to program this estimator is included in the supplementary Web appendix.

The efficient influence curve (EIC) for $\Psi_{CSDE}$, is given by \begin{equation}
\label{eqEEeicfirst}
D_{CSDE}(P) = \frac{D_{SDE}(P)}{\Psi_{FS}(P)} - \frac{\Psi_{SDE}(P)D_{FS}(P)}{\Psi_{FS}^2(P)}, \end{equation}
where $P$ represents $(Q_{W},g_A,g_{Z}, \bar{Q})$, and where
\begin{equation}
\label{eqEEeicmiddle}
\begin{aligned}
D_{SDE}(P) &= \bigg(\frac{g_{1|W,Z}}{g_{1|W}} - \frac{g_{0|W,Z}}{g_{0|W}}\bigg) \frac{\hat{g}_{M|A=0,W} }{g_{M|Z,W}} (Y-\bar{Q}_Y(M,Z,W) )\\
& + \frac{2A-1}{g_{A|W}} (\bar{Q}_M(1,W) - \bar{Q}_M(0,W))(Z-g_Z(1 | A, W))\\
& + (\bar{Q}_{Z}(1,W) - \bar{Q}_{Z}(0,W)) - \Psi_{SDE}
\end{aligned}
\end{equation}
and where
\begin{equation}\label{eqEEeiclast}
D_{FS}(P) = \frac{2A-1}{g_{A|W}}(Z-g_Z(1 | A, W)) + \{(g_{Z}(1,W)- g_{Z}(0,W)) - \Psi_{FS} \}.
\end{equation}


We first solve $D_{SDE}$ to obtain the EE estimate of $\Psi_{SDE}$. We calculate the first component of $D_{SDE}$ as follows, noting that this first component is specifically formulated to respect the exclusion restriction, $Y_{a,z}=Y_z$.
Let $\bar{Q}_Y=E(Y|W,Z,M)$, 
 and let $g_M=P(M=m | Z,W), g_A=P(A=a | W)$, and $g_{A2}=P(A=a | W,Z)$. Recall that $\hat{g}_{M|0,W}$ is treated as known, estimated from the observed data, marginalizing out $Z: \sum\limits^1\limits_{z=0} P(M=m|Z=z,W)P(Z=z|A=0,W)$ \citep{vanderweele2016mediation}. We will use the subscript $n$ to denote estimates throughout. $\bar{Q}_{Y,n}$ can be estimated by predicted values of $Y$ from a regression of $Y$ on $W, Z,$ and $M$. One could use machine learning in model fitting but we will describe estimation in terms of parametric model fitting for simplicity. $g_{A,n}$ can be estimated by predicted probabilities from a logistic regression model of $A$ on $W$. 
  $g_{M,n}$ can be estimated by predicted probabilities from a logistic regression model of $M$ on $Z$ and $W$. 
  $g_{A2}$ can be written $\frac{P(A=a | W)P(Z=z | a, W)}{P(Z=z | W)} = \frac{g_{A|W}g_{Z|A,W}}{P(Z=z | W)} $, where $g_{A,n}$ can be estimated as described above, $g_{Z,n}$ can be estimated from a constrained logistic regression model of $Z$ on $A$ and $W$ to respect the monotonicity assumption, $Z_1-Z_0 \ge 0$, and where an estimate of $P(Z=z | W)$ is obtained by marginalizing out $A: \sum\limits^1\limits_{a=0} P(Z=z|A=a,W)P(A=a | W)$.
 
We now calculate the second component of $D_{SDE}$. To estimate $\bar{Q}_M = E(E(Y|W,Z,M)|W,Z)$, we integrate out $M$ using the data-dependent stochastic intervention on $M$, evaluated at $m|W$, $\hat{g}_{M|0,W}(m|W)$: $\bar{Q}_{M,n} = \sum\limits^1\limits_{m=0}\bar{Q}_{Y,n}(m,Z,W)\hat{g}_{M|0,W}(m|W) $.

Finally, we calculate the third component of $D_{SDE}$. To estimate\\ $\bar{Q}_Z=E(E(E(Y|W,Z,M)|W,Z)|W,A)$, we integrate out $Z$ from\\ $\bar{Q}_{M,n}$: $\bar{Q}_{Z,n} = \sum\limits^1\limits_{z=0}\bar{Q}_{M,n}(z,W)g_{Z,n}$.

The estimate of $\Psi_{SDE}$ is given by solving $D_{SDE}$. The estimate of $\Psi_{FS}$ is given by solving $D_{FS}$, where each component is calculated as described above. The ratio of these two estimates gives the EE estimate of $\Psi_{CSDE}$. The associated variance can be estimated as the sample variance of the EIC, $D_{CSDE}(P)$, which is given in Equation \ref{eqEEeicfirst}. 

\subsubsection{Targeted Minimum Loss-Based Estimator}
We now describe two TMLE approaches to estimate $\Psi_{CSDE}$ using separate estimates for the numerator, $\Psi_{SDE}$, and denominator, $\Psi_{FS}$.\\ 

\noindent \textit{Inefficient TMLE.} The first approach uses a previously developed TMLE for estimating $\Psi_{SDE}$ \citep{rudolph2017robust} and uses the TMLE for an average treatment effect \citep{van2006targeted} for estimating $\Psi_{FS}$. However, using the previously developed TMLE for $\Psi_{SDE}$ neither respects the exclusion restriction or monotonicity constraints on our statistical model, $Y_{a,z}=Y_z$, so we refer to this as an inefficient TMLE.\\

\noindent \textit{Efficient TMLE.} The second approach proposes a novel TMLE for $\Psi_{SDE}$ that respects the exclusion restriction and monotonicity statistical constraints. Additionally, as in the EE estimation approach detailed above, we use the same estimate of the conditional distribution of $Z$ given $A$ and $W$ for both the numerator and denominator, and employ constrained regression in estimating this conditional distribution to enforce the monotonicity statistical constraint. Thus, we refer to this as an efficient TMLE and describe, step-by-step, how to compute this particular TMLE. The R code to program this efficient TMLE estimator is included in the supplementary Web appendix. 

Let $\bar{Q}_Y=E(Y|W,Z,M)$, 
 and let $g_M(m|Z,W)=P(M=m | Z,W), g_A(a|W)=P(A=a | W)$, and $g_{A2}(a|W,Z)=P(A=a | W,Z)$. Recall that $\hat{g}_{M|0,W}$ is treated as known, estimated from the observed data, marginalizing out $Z: \sum\limits^1\limits_{z=0} P(M=m|Z=z,W)P(Z=z|A=0,W)$ \citep{vanderweele2016mediation}. We will use the subscript $n$ to denote estimates throughout. Consider submodel $\{\bar{Q}_{Y,n}(M,Z,W)(\epsilon):\epsilon\}$ defined as:
$logit(\bar{Q}_{Y,n} (\epsilon)(M,Z,W)) = logit(\bar{Q}_{Y,n}(M,Z,W) ) + \epsilon C_Y$, where $C_Y = \bigg(\frac{g_{1|W,Z}}{g_{1|W}} - \frac{g_{0|W,Z}}{g_{0|W}}\bigg) \frac{\hat{g}_{M|A=0,W} }{g_{M|Z,W}}$. This $C_Y$ differs from the $C_Y$ in the previously developed inefficient TMLE for $\Psi_{SDE}$ in that the targeting step does not introduce dependence on $A$ \citep{rudolph2017robust}; instead, the exclusion restriction constraint is preserved. 

These estimators can be calculated as follows. $\bar{Q}_{Y,n}$ can be estimated by predicted values of $Y$ from a regression of $Y$ on $W, Z,$ and $M$. One could use machine learning in model fitting but we will describe estimation in terms of parametric model fitting for simplicity. The components of $C_Y$ can be estimated as follows. $g_{A,n}$ can be estimated by predicted probabilities from a logistic regression model of $A$ on $W$. 
  $g_{M,n}$ can be estimated by predicted probabilities from a logistic regression model of $M$ on $Z$ and $W$. 
  $g_{A2}$ can be written $\frac{P(A=a | W)P(Z=z | a, W)}{P(Z=z | W)} = \frac{g_{A|W}g_{Z|A,W}}{P(Z=z | W)} $, where $g_{A,n}$ can be estimated as described above, $g_{Z,n}$ can be estimated from a constrained logistic regression model of $Z$ on $A$ and $W$, and where an estimate of $P(Z=z | W)$ is obtained by marginalizing out $A: \sum\limits^1\limits_{a=0} P(Z=z|A=a,W)P(A=a | W)$.
 
Let $\epsilon_n$ be the MLE fitted coefficient on $C_Y$ in the logistic regression model of $Y$ on $C_Y$ with $logit \bar{Q}_{Y,n}$ as an offset, using the binary log-likelihood loss function. Alternatively, a non-negative portion of $C_Y$ (e.g., $\frac{\hat{g}_{M|A=0,W} }{g_{M|Z,W} }$) may be moved into the weights and a weighted logistic regression model may be fitted. $Y$ can be bounded to the [0,1] scale as previously recommended \citep{gruber2010targeted}. The updated estimator is given by $ \bar{Q}^*_{Y,n}(M,Z,W) =  \bar{Q}_{Y,n}(\epsilon_n)(M,Z,W)$; noting again that 
 conditional independence with $A$ is preserved.

We next integrate out $M$ using the data-dependent stochastic intervention on $M$, $\hat{g}_{M|0,W}$, to estimate $\bar{Q}_M = E(E(Y|W,Z,M)|W,Z)$: $\bar{Q}^*_{M,n} = \sum\limits^1\limits_{m=0}\bar{Q}^*_{Y,n}(m,Z,W)\hat{g}_{M|0,W}(m|W) $.

The next step is to target $g_{Z,n}$, given above. We denote this targeted $g_{Z,n}$ that is used in the numerator with $g_{Z,n}^{N*}$ to distinguish it from the targeted version that is used in the denominator. Consider submodel $\{g_{Z,n}(\epsilon_1, \epsilon_2):\epsilon_1, \epsilon_2\}$ defined as: $logit g_{Z,n,\epsilon_1, \epsilon_2}(1|W,A) = logit g_{Z,n}(1|W,A) + \epsilon_1 I(A=1)C_Z (\bar{Q}_{M,n}^*(1,W) - \bar{Q}_{M,n}^*(0,W)) + \epsilon_2 I(A=0) C_Z (\bar{Q}_{M,n}^*(1,W) - \bar{Q}_{M,n}^*(0,W)),$ where $C_Z = \frac{1}{g_{A|W}}$. 
Let $\{\epsilon_{1,n}, \epsilon_{2,n}\}$ be the MLE fitted coefficients on $I(A=1)(\bar{Q}^*_{M,n}(1,W) - \bar{Q}^*_{M,n}(0,W))$ and $I(A=0)(\bar{Q}^*_{M,n}(1,W) - \bar{Q}^*_{M,n}(0,W))$ in the weighted logistic regression model of $Z$ with $logit g_{Z,n}$ as an offset, using the binary log-likelihood loss function, and weights $C_Z$. The updated estimator is given by $g^{N*}_{Z,n}=g_{Z,n}(\epsilon_{1,n}, \epsilon_{2,n}).$ 

We can now estimate $\bar{Q}_Z=E(E(E(Y|W,Z,M)|W,Z)|W,A)$ by integrating out $Z$ from $\bar{Q}^*_{M,n}$: $\bar{Q}_{Z,n} = \sum\limits^1\limits_{z=0}\bar{Q}^*_{M,n}(z,W)g^{N*}_{Z,n}(z|A,W)$. 

The estimate of $\Psi_{SDE}$ is given by $Q_{W,n}(\bar{Q}_{Z,n}(1,W) - \bar{Q}_{Z,n}(0,W))$, where $Q_{W,n}$ is the empirical distribution of $W$. It is the empirical mean of the difference in $\bar{Q}_{Z,n}$, setting $a=1$ versus $a=0$. 

The estimate of $\Psi_{FS}$ is given by $Q_{W,n}(g_{Z,n}^{D*}(1|1,W) - g_{Z,n}^{D*}(1|0,W))$. It is the empirical mean of the difference in $g_{Z,n}^{D*}$, setting $a=1$ versus $a=0$. We denote the targeted $g_{Z,n}$ used in the denominator with $g_{Z,n}^{D*}$ to distinguish it from the targeted version used in the numerator, $g_{Z,n}^{N*}$. Targeting $g_{Z,n}$ in the denominator can be done as follows. Consider submodel $\{g_{Z,n}(\epsilon_{D1}, \epsilon_{D2}):\epsilon_{D1}, \epsilon_{D2}\}$ defined as: $logit g_{Z,n,\epsilon_{D1}, \epsilon_{D2}}(1|W,A) = logit g_{Z,n}(1|W,A) + \epsilon_{D1} C_Z I(A=1) + \epsilon_{D2} C_Z I(A=0),$ where $C_Z = \frac{1}{g_{A|W}}$. 
Let $\{\epsilon_{D1,n}, \epsilon_{D2,n}\}$ be the MLE fitted coefficients on $I(A=1)$ and $I(A=0)$ in the weighted logistic regression model of $Z$ with $logit g_{Z,n}$ as an offset, using the binary log-likelihood loss function, and weights $C_Z$. The updated estimator is given by $g^{D*}_{Z,n}=g_{Z,n}(\epsilon_{D1,n}, \epsilon_{D2,n}).$ 

The ratio of these two estimates, $\frac{Q_{W,n}(\bar{Q}_{Z,n}(1,W) - \bar{Q}_{Z,n}(0,W))}{Q_{W,n}(g_{Z,n}^{D*}(1,W) - g_{Z,n}^{D*}(0,W))}$, gives the efficient TMLE estimate of $\Psi_{CSDE}$. The TMLE solves the empirical means of the efficient influence curves (EIC) for $\Psi_{SDE}$ and $\Psi_{FS}$ (Equations \ref{eqEEeicmiddle}-\ref{eqEEeiclast} in the previous section), replacing $P$ with $P_n^*$, where $P_n^*$ represents $(Q_{W,n}, g_{A,n}, g_{Z,n}^{N*}, g_{Z,n}^{D*}, \bar{Q}_{Y,n}^*) $. 


The variance of the TMLE estimator of $\Psi_{CSDE}$ is estimated as the sample variance of $D_{CSDE}(P_n^*)$.


\subsection{TMLE that estimates the CDSE ratio directly}

We now describe a TMLE that targets the CSDE ratio itself. This TMLE is both efficient, because it respects the model constraints, and compatible, because it simultaneously targets the numerator and denominator. We henceforth refer to this as the compatible TMLE and describe, step-by-step, how to compute it. The R code to program this compatible TMLE estimator is included in the supplementary Web appendix. 

Let $\bar{Q}_Y=E(Y|W,Z,M)$, 
 and let $g_M(m|Z,W)=P(M=m | Z,W), g_A(a|W)=P(A=a | W)$, and $g_{A2}(a|W,Z)=P(A=a | W,Z)$. Recall that $\hat{g}_{M|0,W}$ is treated as known, estimated from the observed data, marginalizing out $Z: \sum\limits^1\limits_{z=0} P(M=m|Z=z,W)P(Z=z|A=0,W)$ \citep{vanderweele2016mediation}. We will use the subscript $n$ to denote estimates throughout. Consider submodel $\{\bar{Q}_{Y,n}(M,Z,W)(\epsilon):\epsilon\}$ defined as:
$logit(\bar{Q}_{Y,n} (\epsilon)(M,Z,W)) = logit(\bar{Q}_{Y,n}(M,Z,W) ) + \epsilon C_Y$, where $C_Y = \bigg(\frac{g_{1|W,Z}}{g_{1|W}} - \frac{g_{0|W,Z}}{g_{0|W}}\bigg) \frac{\hat{g}_{M|A=0,W} }{g_{M|Z,W}}$.
Recall that this $C_Y$ preserves the exclusion restriction constraint on our statistical model such that $Y$ is conditionally independent of $A$ given $M,Z,W$.

These estimators can be calculated as follows. $\bar{Q}_{Y,n}$ can be estimated by predicted values of $Y$ from a regression of $Y$ on $W, Z,$ and $M$. One could use machine learning in model fitting but we will describe estimation in terms of parametric model fitting for simplicity. The components of $C_Y$ can be estimated as follows. $g_{A,n}$ can be estimated by predicted probabilities from a logistic regression model of $A$ on $W$. 
  $g_{M,n}$ can be estimated by predicted probabilities from a logistic regression model of $M$ on $Z$ and $W$. 
  $g_{A2}$ can be written $\frac{P(A=a | W)P(Z=z | a, W)}{P(Z=z | W)} = \frac{g_{A|W}g_{Z|A,W}}{P(Z=z | W)} $, where $g_{A,n}$ can be estimated as described above, $g_{Z,n}$ can be estimated from a constrained logistic regression model of $Z$ on $A$ and $W$, and where an estimate of $P(Z=z | W)$ is obtained by marginalizing out $A: \sum\limits^1\limits_{a=0} P(Z=z|A=a,W)P(A=a | W)$.
 
Let $\epsilon_n$ be the MLE fitted coefficient on $C_Y$ in the logistic regression model of $Y$ on $C_Y$ with $logit \bar{Q}_{Y,n}$ as an offset, using the binary log-likelihood loss function. Alternatively, a non-negative portion of $C_Y$ (e.g., $\frac{\hat{g}_{M|A=0,W} }{g_{M|Z,W} }$) may be moved into the weights and a weighted logistic regression model may be fitted. $Y$ can be bounded to the [0,1] scale as previously recommended \citep{gruber2010targeted}. The updated estimator is given by $ \bar{Q}^*_{Y,n}(M,Z,W) =  \bar{Q}_{Y,n}(\epsilon_n)(M,Z,W)$; noting again that conditional independence with $A$ is preserved.

We next integrate out $M$ using the data-dependent stochastic intervention on $M$, $\hat{g}_{M|0,W}$, to estimate $\bar{Q}_M = E(E(Y|W,Z,M)|W,Z)$: $\bar{Q}^*_{M,n} = \sum\limits^1\limits_{m=0}\bar{Q}^*_{Y,n}(m,Z,W)\hat{g}_{M|0,W}(m|W) $.

The next step is to target $g_{Z,n}$, given above, for both the numerator and denominator and in such a way that preserves monotonicity of $A$ on $Z$. Consider submodel $\{g_{Z,n}(\epsilon_1,\epsilon_2,\epsilon_3,\epsilon_4):\epsilon_1, \epsilon_2,\epsilon_3,\epsilon_4\}$ defined as: $logit g_{Z,n,\epsilon_1,\epsilon_2,\epsilon_3,\epsilon_4}(1|W,A) = logit g_{Z,n}(1|W,A) + \epsilon_1 I(A=1)C_Z + \epsilon_2 I(A=0)C_Z + \epsilon_3 I(A=1)C_Z(\bar{Q}_{M,n}^*(1,W) - \bar{Q}_{M,n}^*(0,W)) + \epsilon_4 I(A=0)C_Z(\bar{Q}_{M,n}^*(1,W) - \bar{Q}_{M,n}^*(0,W)),$ where $C_Z = \frac{1}{g_{A|W}}$. 
Let ($\epsilon_{1,n}, \epsilon_{2,n}, \epsilon_{3,n}, \epsilon_{4,n}$) be the MLE fitted coefficients on $I(A=1)$, $I(A=0)$, $I(A=1)(\bar{Q}^*_{M,n}(1,W) - \bar{Q}^*_{M,n}(0,W))$, and $I(A=0)(\bar{Q}^*_{M,n}(1,W) - \bar{Q}^*_{M,n}(0,W))$ in the weighted logistic regression model of $Z$ with $logit g_{Z,n}$ as an offset, using the binary log-likelihood loss function, and weights $C_Z$. The updated estimator is given by $g^*_{Z,n}=g_{Z,n}(\epsilon_{1,n}, \epsilon_{2,n}, \epsilon_{3,n}, \epsilon_{4,n}).$ 

We can now estimate $\bar{Q}_Z=E(E(E(Y|W,Z,M)|W,Z)|W,A)$ by integrating out $Z$ from $\bar{Q}^*_{M,n}$: $\bar{Q}_{Z,n} = \sum\limits^1\limits_{z=0}\bar{Q}^*_{M,n}(z,W)g^*_{Z,n}(z|A,W)$. 

The estimate of $\Psi_{SDE}$ is given by $Q_{W,n}(\bar{Q}_{Z,n}(1,W) - \bar{Q}_{Z,n}(0,W))$, where $Q_{W,n}$ is the empirical distribution of $W$. It is the empirical mean of the difference in $\bar{Q}_{Z,n}$, setting $a=1$ versus $a=0$. The estimate of $\Psi_{FS}$ is given by $Q_{W,n}(g_{Z,n}^*(1|1,W) - g_{Z,n}^*(1|0,W))$. It is the empirical mean of the difference in $g_{Z,n}^*$, setting $a=1$ versus $a=0$. The ratio of these two estimates, $\frac{Q_{W,n}(\bar{Q}_{Z,n}(1,W) - \bar{Q}_{Z,n}(0,W))}{Q_{W,n}(g_{Z,n}^{*}(1,W) - g_{Z,n}^{*}(0,W))}$, gives the TMLE estimate of $\Psi_{CSDE}$. The TMLE solves the empirical mean of the efficient influence curve (EIC) for $\Psi_{CSDE}$ (Equations \ref{eqEEeicfirst}-\ref{eqEEeiclast} in the previous section), replacing $P$ with $P_n^*$, where $P_n^*$ represents $(Q_{W,n}, g_{A,n}, g_{Z,n}^*, \bar{Q}_{Y,n}^*) $. 


The variance of the TMLE estimator of $\Psi_{CSDE}$ is estimated as the sample variance of $D_{CSDE}(P_n^*)$.


\section{Simulation}
\subsection{Overview}
We conduct a simulation study to examine finite sample performance of the IPTW, EE, and TMLE estimators for $\Psi_{CSDE}$ from the two data-generating mechanisms (DGMs) shown in Table \ref{dgmtab}. 
 In the Moving to Opportunity data used for the empirical illustration, we have $O=(W,\Delta, \Delta A, \Delta Z, \Delta M, \Delta Y)$, where $\Delta$ is an indicator of selection into the sample (in the Moving to Opportunity data, one child from each family is selected to participate). This results in a factorized likelihood: \begin{equation}
     \begin{aligned}
   p_0(O) &= p_0(Y|W,Z,M,\Delta=1)p_0(M|W,Z,\Delta=1)p_0(Z|W,A,\Delta=1)\\
   &\times p_0(A|W,\Delta=1)p_0(\Delta=1 |W)p_0(W).
     \end{aligned}
       \end{equation}
     $$ $$ 
  Under the assumptions enumerated in Section 3, our causal quantity of interest, the CSDE, is identified by the statistical parameter, $\Psi_{CSDE} = \Psi_{SDE} / \Psi_{FS},$ where $\Psi_{SDE}$ is identified 
  \begin{equation}
     \begin{aligned}
    \Psi_{SDE} &\equiv E_0(E_0(E_{g^*_{M|0, W}}\{E_0(Y | W,\Delta=1,Z, M) | W, \Delta=1,Z\}| W,\Delta=1, A=1) |W )\\
     & - E_0(E_0(E_{g^*_{M|0, W}}\{E_0(Y | W,\Delta=1,Z, M) | W,\Delta=1,Z \}| W,\Delta=1,A=0)|W),
    \end{aligned}
 \end{equation} and where $\Psi_{FS}$ is identified 
 \begin{equation} \Psi_{FS} \equiv E_0(E_0(Z | W, \Delta=1, A=1)|W) - E_0(E_0(Z | W, \Delta=1, A=0)|W). \end{equation}

 This slight modification to the SCM results in correspondingly slight modifications to the estimators. For the IPTW estimators, the weights are multiplied by the inverse probability of sampling weights $\Delta/\pi$, where $\pi$ represents $P(\Delta=1 | W)$. The EE estimators solve the numerator and denominator of an EIC that is nearly identical to that given in Equations \ref{eqEEeicfirst} - \ref{eqEEeiclast} only now multiplied by $\Delta/\pi$, giving the modified EICs for $\Psi_{SDE}$ and $\Psi_{FS}$: 
\begin{equation}
\begin{aligned}
D_{SDE}(P) &=\frac{\Delta}{\pi} \bigg(\frac{g_{1|W,Z}}{g_{1|W}} - \frac{g_{0|W,Z}}{g_{0|W}}\bigg) \frac{\hat{g}_{M|A=0,W} }{g_{M|Z,W}} (Y-\bar{Q}_Y(M,Z,W) )\\
& + \frac{\Delta}{\pi} \bigg(\frac{2A-1}{g_{A|W}} (\bar{Q}_M(1,W) - \bar{Q}_M(0,W))(Z-g_Z(1 | A, W))\\
& + (\bar{Q}_{Z}(1,W) - \bar{Q}_{Z}(0,W)) \bigg) - \Psi_{SDE}
\end{aligned}
\end{equation}
and where
\begin{equation}
D_{FS}(P) = \frac{\Delta}{\pi}\bigg(\frac{2A-1}{g_{A|W}}(Z-g_Z(1 | A, W)) + (g_{Z}(1,W)- g_{Z}(0,W))\bigg) - \Psi_{FS} .
\end{equation}
The TMLE estimators are now inverse-weighted TMLEs where the clever covariates are multiplied by $\Delta/\pi$. 
 
 Table \ref{dgmtab} uses the same notation as in Section 2, excepting the addition of $\Delta$. The first DGM represents the primary simulation and a moderate-strong instrument scenario. The second DGM represents a weak instrument scenario that may be more likely to result in CSDE estimates that lie outside the bounds of the parameter space when estimated by non-substitution-based estimators.

\begin{table}[!h]
\footnotesize
\centering
\caption{Simulation data-generating mechanisms.}
\label{dgmtab}
\begin{tabular}{| p{10.5cm} | p{2.5cm} | }
  \hline
Moderate-Strong Instrument Simulation&\\
$W_1 \sim Ber(0.5)$ & $P(W_1 = 0.50)$ \\
$W_2 \sim Ber(0.4 + 0.2W_1)$  & $P(W_2 = 0.50)$\\
$\Delta \sim Ber(-1 + log(4)W_1 + log(4)W_2$  & $P(\Delta = 0.58)$\\
$A = \Delta A^*$, where $A^*\sim Ber(0.5)$  & $P(A = 0.50)$ \\
$Z = \Delta Z^*$, where $Z^*\sim Ber(log(4)A - log(2)W_2) $ & $P(Z = 0.58)$\\
$M = \Delta M^*$, where $M^*\sim Ber(-log(3) + log(10)Z - log(1.4)W_2)$ & $P(M = 0.52)$\\
$Y = \Delta Y^*$, where $Y^*\sim Ber(log(1.2) + log(3)Z + log(3)M - log(1.2)W_2 + log(1.2)ZW_2)$ & $P(Y = 0.76)$\\
   \hline
Weak Instrument Simulation&\\
$Z = \Delta Z^*$, where $Z^*\sim Ber(0.005 + 0.1 A + 0.5W_2) $ & $P(Z = 0.31)$\\
   \hline
\end{tabular}
\end{table}

We compare performance of our three estimators. 
 We show estimator performance in terms of absolute bias, percent bias, closeness to the efficiency bound (mean estimator standard error (SE) $\times$ the square root of the number of observations), 95\% confidence interval (CI) coverage, and mean squared error (MSE) across 1,000 simulations for sample sizes of N=5,000, N=500, and N=100. In addition, we consider 1) correct specification of all models, 2) misspecification of the $Y$ model that included a term for $Z$ only, 3) misspecification of the $M$ model that included a term for $W$ only, 4) misspecification of the $M$ and $Y$ models, 5) misspecification of the $Z$ model that included a term for $A$ only, and 6) misspecification of the $Z$ and $Y$ models. 

\subsection{Results}
Table \ref{restab} gives results under the moderate-strong instrument simulation scenario using correct model specification for $\Psi_{CSDE}$, comparing the TMLE , IPTW, and EE estimators. We see that the TMLE, IPTW, and EE estimators are consistent when all models are correctly specified and sample sizes are large (N=5,000), showing biases of less than 1\% in the case of the TMLE and EE estimators and just over 1\% in the case of the IPTW estimator. Bias increases for the IPTW estimator under the smaller sample sizes of N=500 and N=100 to 4\% and 42\% respectively, indicating that even under correct model specification, this estimator is challenged in finite samples. In contrast, the TMLE and EE estimators continue to perform well when sample size decreases to N=500 and N=100, although the efficient TMLE that targets the numerator and denominator separately shows a bias of 13\% under N=100. The compatible TMLE and EE estimators perform similarly and close to the efficiency bound for all sample sizes, though efficiency decreases slightly with decreasing sample size. 95\% CI coverage for both is close to 95\% for N=5,000 and N=500 and is reduced slightly for N=100. 95\% CI coverage is conservative for the IPTW estimator---around 99\% for all sample sizes. 

\begin{table}[h!]
\footnotesize
\centering
\caption{ \footnotesize Simulation results comparing estimators of $\Psi_{CSDE}$ under  correct model specification for various sample sizes. 1,000 simulations. Estimation methods compared include IPTW, EE, efficient TMLE, and compatible TMLE. Bias and MSE values are averages across the simulations. The estimator standard error $\times \sqrt{n}$ should be compared to the efficiency bound, which is 1.10.}
\label{restab}
\begin{tabular}{|p{3.3cm}| p{1.5cm}  p{1.5cm} p{2cm}  p{2.2cm}  p{1.5cm}  | }
  \hline
Estimand & Bias & \%Bias & SE$\times \sqrt{n}$ & 95\%CI Cov & MSE \\   \hline 
\multicolumn{6}{| c | }{All correctly specified} \\ \hline
N=5000&&&&&\\
TMLE, compatible  & 0.000 &0.07 &1.11& 94.90 & 0.000\\ 
TMLE, efficient & 0.000& 0.07 & 1.11 & 94.90 & 0.000\\
IPTW &0.003 &  1.45  &6.53& 98.70 &  0.005\\ 
EE &0.000 &0.12 & 1.11 & 94.50 & 0.000\\
\hline
N=500&&&&&\\
TMLE, compatible & -0.001 &-0.47 &  1.11& 94.90&  0.002 \\
TMLE, efficient &-0.001 & -0.47 &  1.11 & 95.00 &  0.002\\
IPTW &-0.009 & -4.08 &  6.89 & 98.40 &  0.051 \\
EE& -0.002 & -0.75 &  1.11 &  95.50 &  0.002 \\
\hline
N=100&&&&&\\
TMLE, compatible& -0.009&  -3.96&  1.14& 90.64&  0.014 \\
TMLE, efficient & 0.028& 13.43 &  1.17 & 86.50 &  0.045\\
IPTW&-0.093 & -42.39 &   21.71 &  99.20 &   1.484 \\
EE &-0.005 & -2.15 &  1.12 & 93.30 &  0.013\\
\hline
\end{tabular}
\end{table}

Table \ref{restab2} gives simulation results under various model misspecifications with large sample size of N=5,000. The IPTW estimator is consistent if the $A$ and $M$ models are correctly specified. Deriving the robustness properties for the EE and TMLE estimators from the EIC, under large sample size, one of three scenarios is required for estimates of $\Psi_{SDE}$ to be consistent: 1) the $g_{A|W}$, $g_{Z|A,W}$, and $g_{M|Z,W}$ 
 models need to be correctly specified, or 2) the $g_{Z|A,W}$ and $\bar{Q}_Y$ 
 models need to be correctly specified, or 3) the $g_{A|W}$, $g_{M|Z,W}$, and $\bar{Q}_Y$ 
 models need to be correctly specified. For estimates of $\Psi_{FS}$ to be consistent, either the $g_{A|W}$ or $g_{Z|A,W}$ 
 model needs to be correctly specified. Thus, for the EE and TMLE estimators, robustness requirements for the denominator are subsumed in the robustness requirements for the numerator. In the simulation results that follow, we note that $A$ is randomly assigned in the DGM we consider, aligned with its role as an instrument and with our motivating example.

As expected from each estimator's robustness properties, we see that all estimators are consistent under misspecification of the $Y$ model, with performance equivalent to performance under correct model specifications and N=5,000. Also as expected, under misspecification of the $M$ model, the IPTW estimator is no longer consistent with 11\% bias, but the TMLE and EE estimators remain consistent. When both the $M$ and $Y$ models are misspecified, all three estimators are inconsistent with biases ranging from 11\% for IPTW to 45\% for TMLE and EE, and 95\% CI coverage of the TMLE and EE estimators is reduced to 0\%.  The compatible TMLE, EE, and IPTW 
 estimators are consistent under misspecification of the $Z$ model, but the IC-based inference for the TMLE and EE estimators is no longer valid, resulting in under coverage. Coverage improves when bootstrapping is used for inference (Table \ref{restab2}). In this scenario, the efficient TMLE demonstrates slight bias (6\%), possibly because $g_{Z|A,W}$ is targeted separately in the numerator and denominator, resulting in incompatibility. 
  Misspecification of both the $Z$ and $Y$ models results not only in invalid inference for the EE and TMLE estimators but also in inconsistent estimates. We note that for the two scenarios where the $Z$ model was misspecified the true DGM was changed to $Z^* \sim Ber(log(4)A -log(40)W)$ to make misspecifying the $Z$ model meaningful.

\begin{table}[h!]
\footnotesize
\centering
\caption{\footnotesize Simulation results comparing estimators of $\Psi_{CSDE}$ under various model misspecifications 
 for sample size N=5,000. 1,000 simulations. Estimation methods compared include IPTW, EE, efficient TMLE, and compatible TMLE. 
  Bias and MSE values are averages across the simulations. The estimator standard error $\times \sqrt{n}$ should be compared to the efficiency bound, which is 1.10
. 95\% CI Coverage as determined by bootstrapping is denoted in parentheses for scenarios in which the $Z$ model is misspecified.}
\label{restab2}
\begin{tabular}{|p{3.2cm}| p{1.5cm}  p{1.5cm} p{2cm}  p{2.5cm}  p{1.2cm}  | }
  \hline
Estimand & Bias & \%Bias & SE$\times \sqrt{n}$ & 95\%CI Cov & MSE \\   \hline 
\multicolumn{6}{| c | }{M model misspecified, N=5,000} \\ \hline
TMLE, compatible &0.000 &  0.02&  1.05& 94.30&  0.000\\ 
TMLE, efficient & 0.000&0.02 & 1.05 & 94.30 & 0.000\\
IPTW & -0.024& -11.28 &   5.32 &  99.40&   0.003\\ 
EE &0.000 & 0.08 &1.05 & 94.20 & 0.000\\

\hline
\multicolumn{6}{| c | }{Y model misspecified, N=5,000} \\ \hline
TMLE, compatible &0.000 &  0.09&  1.14& 95.60&  0.000\\ 
TMLE, efficient & 0.000 & 0.09 & 1.14 & 95.60 &0.000 \\
IPTW&0.003 & 1.45&  6.53& 98.70&  0.005\\
EE&0.000 & 0.06 &  1.19 & 96.10 & 0.000\\
\hline
\multicolumn{6}{| c | }{M and Y models misspecified, N=5,000} \\ \hline
TMLE, compatible &0.096 &  44.90&  1.06& 0.00&  0.009\\
TMLE, efficient & 0.096& 44.90 &  1.06 & 0.00& 0.009\\
IPTW&-0.024& -11.28 &   5.32&  99.40   &0.003\\
EE&  0.095 & 44.74 &  1.06 &  0.00 &  0.009\\ \hline
\multicolumn{6}{| c | }{Z models misspecified, N=5,000} \\ \hline
TMLE, compatible & 0.001 & 0.28 & 1.16& 87.50 (92.90) &  0.000\\
TMLE, efficient &0.012 & 6.28 & 1.38 & 83.10 (92.60) & 0.001\\
IPTW &0.003 &  1.45  &6.53& 98.70 &  0.005\\ 
EE & 0.001 & 0.26 & 1.16 & 88.30 (92.80) & 0.000\\
\hline
\multicolumn{6}{| c | }{Z and Y models misspecified, N=5,000} \\ \hline
TMLE, compatible & 0.066 & 33.64& 1.16 &  2.60 (47.00) & 0.005\\
TMLE, efficient &0.041& 20.72 & 1.42& 45.70 (46.90)& 0.002\\
IPTW &0.003 &  1.45  &6.53& 98.70 &  0.005\\
EE & 
 0.071 & 36.09&  1.23 &  2.10 (38.30)&  0.005\\
\hline
\end{tabular}
\end{table}

Table \ref{restab3} gives results under the weak instrument simulation scenario using correct model specification, comparing the IPTW, EE, efficient TMLE, and compatible TMLE estimators. Finite sample performance is challenged in this scenario---we see all estimators having larger biases and worse efficiency for a given sample size than in the stronger instrument scenario. Performances of the IPTW estimator is particularly affected. Under sample size N=500 and correct model specification, the IPTW estimator is 27\% biased compared to 3\% bias of the other estimators.
 With sample size N=100 under this weak instrument scenario, all estimators perform poorly, with the IPTW estimator displaying particularly egregious performance. 

In part, the poor efficiency of the IPTW estimator is 
due to very small or even negative denominator estimates in this weak instrument scenario, which results in CSDE estimates lying outside of the parameter space. Indeed, we see that 
18\% and 53\% of the IPTW estimates were out of the bounds of the parameter space for the N=500 and N=100 sample sizes, respectively. In contrast, the EE and TMLE estimates largely stay within the parameter space for N=500.  For the smallest sample size of N=100, about 3-4\% lie outside of the parameter space for the EE and TMLE estimates.

\begin{table}[h!]
\footnotesize
\centering
\caption{ \footnotesize Simulation results comparing the 
 efficient and compatible TMLE, IPTW, and EE estimators of $\Psi_{CSDE}$ under the weak instrument simulation scenario and correct model specification for various sample sizes. 1,000 simulations. 
 Bias and MSE values are averages across the simulations. The estimator standard error $\times \sqrt{n}$ should be compared to the efficiency bound, which is 1.13.}
\label{restab3}
\begin{tabular}{|p{2cm}| p{1.85cm}  p{1.85cm} p{1.5cm}  p{1.9cm}  p{1.5cm}  p{1.5cm}  | }
  \hline
Estimand & Bias & \%Bias & SE$\times \sqrt{n}$ & 95\%CI Cov & MSE & \% Out of Bounds\\   \hline 
N=5000&&&&&&\\
TMLE compatible& -0.000  &-0.08&  1.13& 73.50&  0.001 & 0.00\\ 
TMLE, efficient & -0.000& -0.07 &  1.13 & 73.50 &  0.001 & 0.00
\\
IPTW & 0.016 & 8.14 & 19.29 & 98.70 & 0.045 &0.10\\
EE &
-0.000 & -0.07 & 1.13 & 74.40 &  0.001 & 0.00\\
\hline
N=500&&&&&&\\
TMLE compatible &0.008 &  3.53&  1.38& 74.60&   0.010& 0.10 \\
TMLE, efficient & 
0.008 &  3.55 &  1.38 & 74.50 &  0.010 & 0.10\\
IPTW & 0.057 & 27.05 & 28.51& 99.90&  0.964 & 18.80\\
EE 
&0.008  &3.69 &  1.40 & 75.30 &  0.010 & 0.10\\
\hline
N=100&&&&&&\\
TMLE compatible & 0.048 &22.19 &  344.04& 86.49&  3.302 &4.10\\
TMLE, efficient & 0.112&  50.95 & 1226.43&  88.51 &   3.550
&4.69\\
IPTW &-1.14$\times 10^{12}$ &-5.11$\times 10^{14}$ & 5.37$\times 10^{28}$ & 100.00&  1.24$\times 10^{27}$ & 53.10\\
EE & 0.042 &   19.40& 53.66&  88.50 & 0.740 & 3.30\\
\hline
\end{tabular}
\end{table}

\section{Empirical Illustration}
\subsection{Overview and set-up}
We now apply our proposed TMLE estimator to MTO: a longitudinal, randomized trial where families living in public housing were randomized to receive a Section 8 housing voucher that they could then use to move out of public housing \citep{kling2007experimental}. In this example, the CSDE is the direct effect of using the housing voucher to move out of public housing on adolescent substance use outcomes, not mediated by aspects of parental well-being, among those who comply with the intervention. 

The instrument, $A$, is defined as randomization to receive a Section 8 housing voucher that one can then use to rent on the private market. The exposure, $Z$, is defined as adherence to the intervention---using the housing voucher, if one received it, to move out of public housing. We examined direct effects not operating through each of five mediators, $M$, all measured at an interim assessment that occurred 4-7 years after the baseline assessment: 1) parental employment; 2) parental anxiety, defined as feeling worried, tense or anxious most of the time or worrying much more than others in his/her situation for at least one month during the past year; 3) parental depression, consistent with \textit{DSM-IV} diagnostic criteria, as measured by the CIDI-SF instrument \citep{kessler1998world}; 4) parental distress, as measured by the Kessler Psychological Distress Scale (K6) \citep{kessler2002short}; and 5) parental warmth towards the adolescent, as measured by direct observation, consisting of nine items. The first three mediators were binary. The last two were indices bounded $[0, 1]$. We examined three adolescent substance use outcomes, $Y$, which were also measured at the interim assessment: past-month cigarette use, past-month marijuana use, and past-month problematic drug use. Problematic drug use was defined as using hard drugs or using marijuana before school or work in the past month. We used a high-dimensional vector of covariates measured at baseline, $W$, that included social-demographic information for the adolescent and his/her family, information on the adolescent's behavior and learning while a child, neighborhood characteristics, and reasons for the family's participation in MTO. Definitions of $W$, $A$, $Z$, and $Y$ align with previous work estimating direct and indirect effects of $A$ on $Y$ in the MTO study \citep{rudolph2017mediation}. These variables follow the same structural causal model as detailed in Section \ref{notationsection}.

Aligned with a prior analysis estimating direct effects in MTO \citep{rudolph2017mediation}, our sample includes adolescents participating in MTO who were 12-17 years old at the interim assessment. We exclude the Baltimore site, as Section 8 voucher receipt did not increase a family's likelihood of moving to a low-poverty neighborhood, which differs from other sites and from the intention of the intervention. We conducted analyses stratified by gender, as previous work documented qualitatively and quantitatively different intervention effects between girls and boys \citep{orr2003moving,clampet2011moving}. We combine sites with similar intervention effects, as has been done previously \citep{rudolph2017mediation}. Lastly, we restrict to those with nonmissing mediator and outcome data. Multiple imputation by chained equations \citep{buuren2011mice} was used to create 30 imputed datasets to address missing covariate values (none had more than 5\% missing). The University of California, Davis, determined this analysis of deidentified data to be non-human subjects research.

\subsection{Results}

Total complier average causal effects (CACEs) \citep{angrist1996identification} (also called treatment-on-treated effects \citep{orr2003moving}) are shown in Web Figures 1-3 (see supplementary Web appendix). This is the effect of $Z$ on $Y$ among compliers, using randomization of the instrument $A$. In other words, it is the total effect of moving with the voucher out of public housing on the outcome, among those who comply with the intervention. Moving with the voucher out of public housing increased risk of cigarette use among boys by 8\% (RD: 0.08, 95\%CI: -0.00, 0.17) and reduced risk of marijuana use among girls by 7\% (RD: -0.07, 95\%CI: -0.13, -0.01). This aligns with previous work finding that the intervention generally improved health and risk behavior outcomes among girls but had negative impacts in terms of these same types of outcomes for boys \citep{orr2003moving,clampet2011moving}.

We next estimated the first-stage, data-dependent, stochastic effect of $A$ on each mediator, $M$: $E(\hat{g}_{M|1,W} - \hat{g}_{M|0,W})$ that are used in each of CSDE estimators. These first-stage effects are shown in Table \ref{stochinttab}. Across outcome samples and genders, we see that being randomized to receive a Section 8 voucher increases parental employment and anxiety but decreases parental depression. Effects of voucher receipt on distress and warmth were mixed or null.

\begin{table}[h!]
\centering
\footnotesize
\caption{Risk differences of the effect of voucher receipt on the mediator by outcome sample (marginal effects, adjusting for baseline covariates and adherence, $Z$).}
\label{stochinttab}
\footnotesize
\begin{tabular}{| l | c | c |}
  \hline
Mediator & Boys & Girls  \\ 
 & RD (95\% CI) & RD (95\% CI) \\ 
  \hline
Cigarette Use Sample&&\\
Parental employment & 0.058 (0.047, 0.070) & 0.026 (0.015, 0.037 ) \\ 
Parental anxiety & 0.036 (0.028, 0.044)& 0.041 (0.039, 0.043) \\ 
Parental depression & -0.004 (-0.007, -0.001) & -0.004 (-0.006, -0.001) \\ 
Parental distress & 0.004 (-0.001, 0.009) & 0.013 (0.009, 0.017)\\ 
Parental warmth & -0.006 (-0.032, 0.019)  & -0.005 (-0.028, 0.019)\\ 
Marijuana Use Sample&&\\
Parental employment& 0.079 (0.067, 0.092)  & 0.027 (0.015, 0.038) \\ 
Parental anxiety  & 0.011 (0.002, 0.021) & 0.042 (0.041, 0.043)\\ 
Parental depression& -0.024 (-0.026, -0.021) & -0.003 (-0.005, -0.001) \\ 
Parental distress& -0.021 (-0.027, -0.015)& 0.012 (0.008, 0.016) \\ 
Parental warmth &  0.005 (-0.023, 0.033) & -0.001 (-0.025, 0.023)\\ 
Problematic Drug Use Sample &&\\
Parental employment& 0.061 (0.052, 0.070)  & 0.052 (0.041, 0.063)   \\ 
Parental anxiety& 0.039 (0.031, 0.047) &  0.050 (0.045, 0.056) \\ 
Parental depression& -0.006 (-0.009, -0.003)& -0.011 (-0.016, -0.005)  \\ 
Parental distress& 0.004 (-0.001, 0.009)   & 0.016 (0.009, 0.022)\\ 
Parental warmth &-0.005 (-0.027, 0.017)  & -0.011 (-0.040, 0.017) \\ 
   \hline
\end{tabular}
\end{table}

\begin{figure}[h!]
\centerline{\includegraphics[width=6.5in]{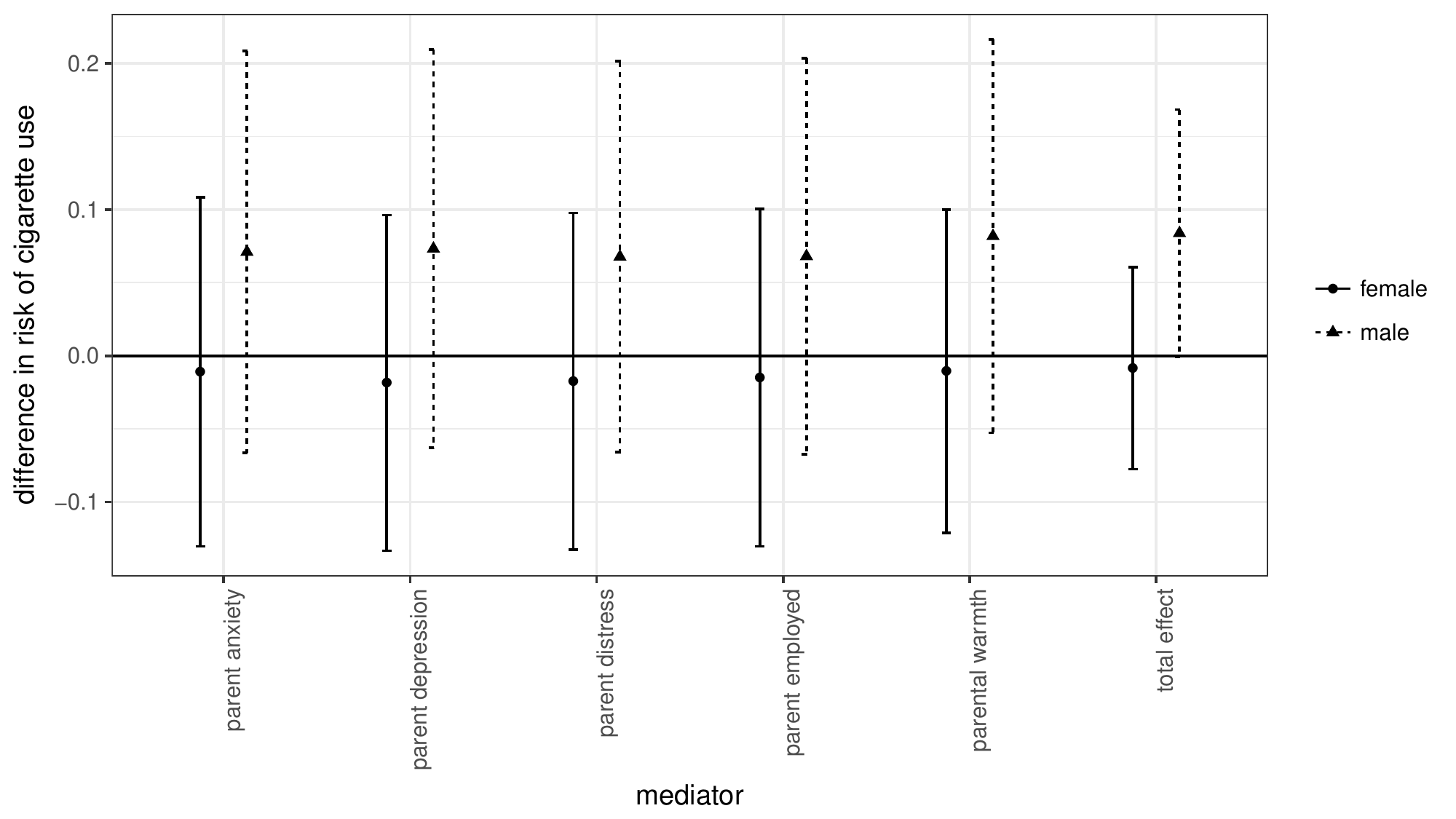}}
\caption{Data-dependent complier stochastic direct effect estimates and 95\% confidence intervals on past-month cigarette use by mediator. Data from the Moving to Opportunity experiment, interim follow up. 
}\label{cigplot}

\end{figure}

\begin{figure}[h!]
\centerline{\includegraphics[width=6.5in]{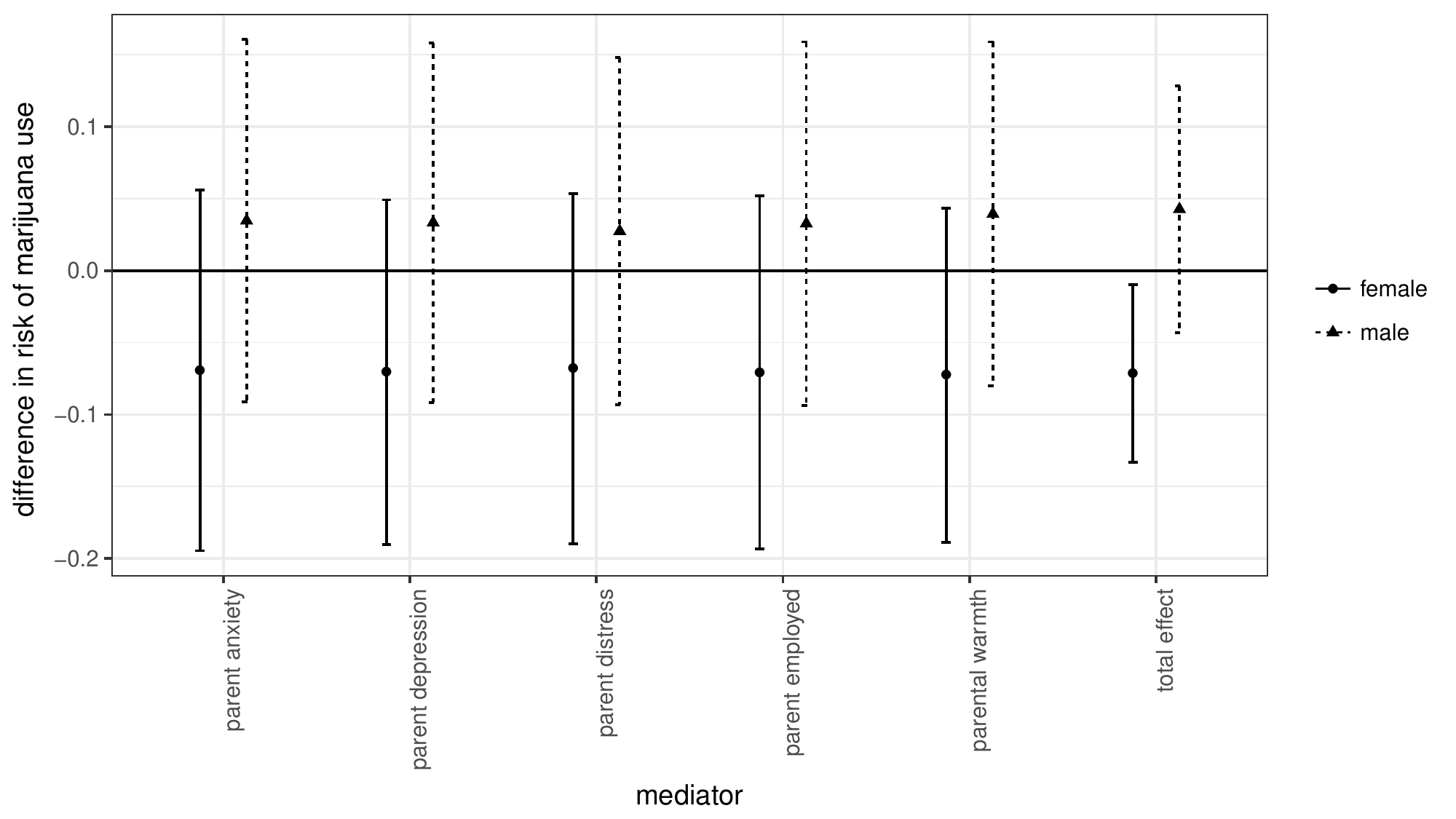}}
\caption{Data-dependent complier stochastic direct effect estimates and 95\% confidence intervals on past-month marijuana use by mediator. Data from the Moving to Opportunity experiment, interim follow up.
}\label{potplot}
\end{figure}

\begin{figure}[h!]
\centerline{\includegraphics[width=6.5in]{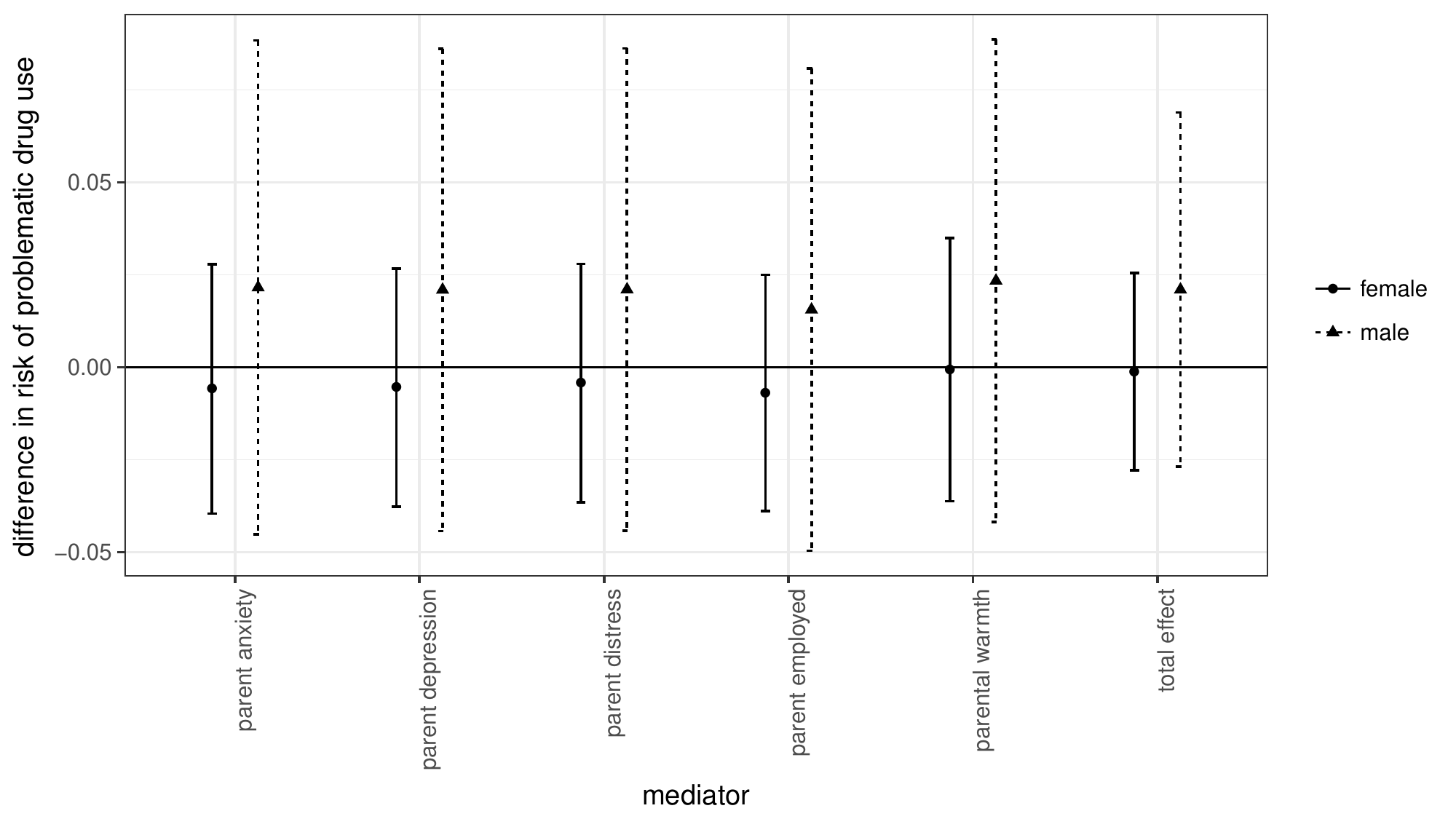}}
\caption{Data-dependent complier stochastic direct effect estimates and 95\% confidence intervals on past-month problematic drug use by mediator. Data from the Moving to Opportunity experiment, interim follow up. 
}\label{prbdrugplot}
\end{figure}

The TMLE estimates of the $\Psi_{CSDE}$s by outcome sample, gender, and mediator are shown in Figures \ref{cigplot} - \ref{prbdrugplot}. The estimates are similar across mediators. Each estimate of the CSDE is similar to its corresponding SDE estimate except for a wider confidence interval, which is anticipated since we are conditioning on compliers (e.g., CACEs have wider confidence intervals than average causal effects). Similar point estimates coupled with wider confidence intervals result in universally null CSDEs. 

Lastly, we compare the CSDE estimates in Figures \ref{cigplot} - \ref{prbdrugplot} with the stochastic direct and indirect effect (SDEs and SIEs) estimates (see supplementary Web appendix Figures 1-3). The SDE is the direct effect of $A$ on $Y$ not through $M$ and the SIE is the indirect effect of $A$ on $Y$ through $M$. The total intent-to-treat average treatment effects (the total effect of $A$ on each $Y$) are included in the figures. Such further examination may be of interest, because previous research has shown that significant indirect effects may be present with null total or direct effects \citep{imai2010identification}. However, all SDE and SIE effect estimates are null. 

Together, these results suggest that none of the five parental well-being variables tested are on the causal pathway from Section 8 voucher receipt to subsequent adolescent substance use. The first-stage results (Table \ref{stochinttab}) provided evidence against mediation by parental distress, or warmth. The CSDE and SDE/SIE estimates then provided evidence against mediation by the remaining variables of parental employment, anxiety, or depression.

\section{Conclusion}
In this paper, we identified the IV direct effect of exposure, $Z$, on outcome $Y$ not operating through mediator, $M$---what we call the complier stochastic direct effect (CSDE). We detailed three estimators to estimate such effects: a ratio of inverse-probability of treatment-weighted estimators (IPTW), a ratio of estimating equation estimators (EE), a ratio of targeted minimum loss-based estimators (TMLE), and a TMLE that targets the CSDE directly. These estimators would be applicable for a variety of study designs, including 1) randomized encouragement trials, like the MTO housing voucher experiment we consider as an illustrative example, 2) treatment discontinuities, such when a policy or practice changes abruptly either over time or at a certain value, and 3) Mendelian randomization \citep{baiocchi2014instrumental}. To facilitate implementation of our proposed estimators, we include step-by-step instructions in the main text and commented R code in the supplementary Web appendix.

Estimators of the CSDE will be challenged by the finite sample sizes encountered in real-world data. Both mediation estimators and IV estimators are less efficient than their total average treatment effect counterparts. Because estimators of the CSDE combine both mediation and IV components, these estimators will likewise have efficiency challenges, particularly in finite samples. Thus, it is important to choose an estimator that is more robust to finite sample bias.

We found the IPTW estimator to be the most sensitive to finite sample bias, resulting in bias of over 40\% even when all models were correctly specified in a sample size of N=100 (Table \ref{restab}). In contrast, the EE estimator and compatible TMLE estimator 
 were far less sensitive to finite samples, demonstrating slight losses of efficiency in sample sizes of N=100. 

The EE and TMLE estimators also have advantages over the IPTW estimator in terms of efficiency and reduced reliance on correct parametric model specification due to 1) being robust to certain combinations of model misspecifications and 2) having theory-based inference when incorporating data-adaptive methods like machine learning into model fitting. In addition, 
 the 
 compatible TMLE estimator solves the efficient influence equation of the CSDE ratio, with the targeting being done in such a way that it is compatible across the numerator and denominator. This compatibility improves the TMLE estimator's performance particularly in the weaker instrument scenario, as was shown in Table \ref{restab3} comparing the compatible and efficient TMLEs. A previous TMLE for the complier average total effect (as opposed to the complier direct effect) used a separate TMLE for each of the numerator and denominator, so did not have the advantage of this compatibility \citep{rudolph2016robust}.

However, the estimators we propose are limited in that they use a data-dependent stochastic intervention on $M$, $\hat{g}_{M|a,W}$, which assumes that the stochastic draw is from a known distribution of $M | a, W$, estimated from the observed data. It would be significantly more complex to solve the compatible EIC for the non-data-dependent version, however, we plan to complete such an extension. 

Perhaps the most significant limitation is that we were unable to identify a corresponding complier stochastic indirect effect without additional restrictive assumptions. This limitation is corroborated by recent work by \citep{frolich2017direct} where a similar IV indirect effect could only be identified by assuming two distinct instruments, one for $Z$ and one for $M$ that themselves were conditionally independent, $A_1 \independent A_2 |W$. 



\bibliographystyle{rss}
\bibliography{bibfile}
\appendix

\section{Identification proof}
\noindent Proof.
\begin{eqnarray*}
\Psi_{CSDE}(P) &\equiv& \{E_0(E_0(E_{g^*_{M|0, W}}\{E_0(Y | W,Z, M) | W,Z \} W,A=1) |W) \\
&& - E_0(E_0(E_{g^*_{M|0, W}}\{E_0(Y | W,Z, M) | W,Z \} W,A=0) |W)\} \\
&& / \{E_0(E_0(Z | W, A=1)|W) - E_0(E_0(Z | W, A=0)|W)\}\\
\text{By assumption 1,}&& P(Z=z\mid W,A=a)=P(Z_a=z\mid W), \text{ so}\\
&\equiv& \{E_0(E_0( E_{g^*_{M|0, W}}\{E_0(Y_{g^*_{M|0,W}}\mid W,Z) \}\mid W, Z_1)\mid W)\\ 
&& - E_0(E_0( E_{g^*_{M|0, W}}\{E_0(Y_{g^*_{M|0,W}}\mid W,Z) \}\mid W, Z_0)\mid W) \} \\
&& / [E_0\{E_0(Z_1 | W) - E_0(Z_0 | W) \}]\\
&\equiv& \{E_0(E_0(Y_{1,g^*_{M|0,W}} | W) - E_0(Y_{0,g^*_{M|0,W}} | W)) \} \\
&& /\{E_0(E_0(Z_1 | W) - E_0(Z_0 | W)) \}\\
&\equiv& \frac{E_0(Y_{1,g^*_{M|0,W}} - Y_{0,g^*_{M|0,W}}) }{E_0(Z_1 - Z_0)} \\
&\equiv& \{E_0(Y_{1,g^*_{M|0,W}} - Y_{0,g^*_{M|0,W}} | Z_1 - Z_0=1)P(Z_1 - Z_0=1) \\
&& + E_0(Y_{1,g^*_{M|0,W}} - Y_{0,g^*_{M|0,W}} | Z_1 - Z_0=0)P(Z_1 - Z_0=0)\\
&& + E_0(Y_{1,g^*_{M|0,W}} - Y_{0,g^*_{M|0,W}} | Z_1 - Z_0=-1)P(Z_1 - Z_0=-1)\} \\
&& / E_0(Z_1 - Z_0)\\
\text{By assumption 2,}&& \\ 
&\equiv& \{E_0(Y_{1,g^*_{M|0,W}} - Y_{0,g^*_{M|0,W}} | Z_1 - Z_0=1)P(Z_1 - Z_0=1) \\
&& + E_0(Y_{1,g^*_{M|0,W}} - Y_{0,g^*_{M|0,W}} | Z_1 - Z_0=-1)P(Z_1 - Z_0=-1)\} \\
&& / E_0(Z_1 - Z_0)\\
\text{By assumption 3,}&& \\ 
&\equiv& \{E_0(Y_{1,g^*_{M|0,W}} - Y_{0,g^*_{M|0,W}} | Z_1 - Z_0=1)P(Z_1 - Z_0=1) \} \\
&& / E_0(Z_1 - Z_0)\\
&& Z_1 - Z_0 \in \{0,1\}, \text{ so } \\
&\equiv& \{E_0(Y_{1,g^*_{M|0,W}} - Y_{0,g^*_{M|0,W}} | Z_1 - Z_0=1)E(Z_1 - Z_0) \} \\
&& / E_0(Z_1 - Z_0)\\
&\equiv& E_0(Y_{1,g^*_{M|0,W}} - Y_{0,g^*_{M|0,W}} | Z_1 - Z_0=1)
\end{eqnarray*}
\noindent { } By assumptions 4 and 5, we have that $\Psi_{CSDE}$ is defined. $\Box$

\section{Estimator modifications when there is also a direct effect of $A$ on $M$}
The complier stochastic direct effect estimand and estimation approaches we consider also work in the scenario where $M$ may depend on $A$ conditional on $Z$: $M=f(W, A, Z, U_M)$. We describe differences in the estimator details for such a scenario here. In this alternative scenario, $A$ is not an instrument for the total effect of $Z$ on $Y$, and the estimation approach suggested by \cite{joffe2008extended} would also be appropriate.
 
The true distribution $P_0$ of $O$ can be factorized as $$ P_0(O) = P_0(Y | W,Z,M)P_0(M|W,A,Z)P_0(Z|W,A)P_0(A|W)P_0(W).$$

\subsection{Inverse Probability of Treatment Weighted Estimator}

The inverse probability of treatment weights for estimating $\Psi_{SDE}$ 
 are \begin{equation} IPTW_{SDE}=\frac{(2A-1)\hat{g}_{M|0,W}}{g_{A|W}g_{M|Z,A,W} }. \end{equation}
 Let $g_{A,n}$ and $g_{M,n}$ be estimators of $g_{A|W}=P(A=a | W)$ and $g_{M|Z,A,W}=P(M=m | Z,A,W)$, respectively. $g_{A,n}$ can be estimated by predicted probabilities from a logistic regression model of $A$ on $W$. One could use machine learning in model fitting but we will describe estimation in terms of parametric model fitting for simplicity. $g_{M,n}$ can be estimated by 
  predicted probabilities from a logistic regression model of $M$ on $W, A, Z$. $\hat{g}_{M|0,W}$ is treated as known, estimated from the observed data, marginalizing out $Z: \sum\limits^1\limits_{z=0} P(M=m|Z=z,A=0,W)P(Z=z|A=0,W)$ \citep{vanderweele2016mediation}. The IPTW estimate of $\Psi_{SDE}$ is the empirical mean of outcome, $Y$, weighted by an estimate of $IPTW_{SDE}$.

The inverse probability of treatment weights for estimating $\Psi_{FS}$ 
 are as written in the main text. 

The associated variance can be estimated as the sample variance of the estimator's influence curve (IC), which is 
\begin{equation}
D_{IPTW}(P) = \frac{D_{IPTW_{SDE}}(P)}{\Psi_{FS}(P)} - \frac{\Psi_{SDE}(P)D_{IPTW_{FS}}(P)}{\Psi_{FS}^2(P)}, \end{equation}
and where 
\begin{equation}
D_{IPTW_{SDE}}(P) = \frac{(2A-1)\hat{g}_{M|0,W}}{g_{A|W}g_{M|Z,A,W}} Y - \Psi_{SDE}
\end{equation}
and where
\begin{equation}
D_{IPTW_{FS}}(P) = \frac{2A-1}{g_{A|W}}Z - \Psi_{FS}.
\end{equation}

\subsection{Estimating Equation Estimator}

This estimator solves the efficient influence curve (EIC) for $\Psi_{CSDE}$, which is given by \begin{equation}
\label{eqEEeic}
D_{CSDE}(P)(Q_{W},g_A,g_{Z}, \bar{Q}) = \frac{D_{SDE}(P)}{\Psi_{FS}(P)} - \frac{\Psi_{SDE}(P)D_{FS}(P)}{\Psi_{FS}^2(P)}, \end{equation}
where 
\begin{equation}
\begin{aligned}
D_{SDE}(P) &= \bigg(\frac{g_{1|W,Z,M}}{g_{1|W}} - \frac{g_{0|W,Z,M}}{g_{0|W}}\bigg) \frac{\hat{g}_{M|A=0,W} }{g_{M|Z,A,W}} (Y-\bar{Q}_Y(M,Z,W) )\\
& + \frac{2A-1}{g_{A|W}} (\bar{Q}_M(Z=1,W) - \bar{Q}_M(Z=0,W))(Z-g_Z(1 | A, W))\\
& + (\bar{Q}_{Z}(A=1,W) - \bar{Q}_{Z}(A=0,W)) - \Psi_{SDE}
\end{aligned}
\end{equation}
and where
\begin{equation}
D_{FS}(P) = \frac{2A-1}{g_{A|W}}(Z-g_Z(1 | A, W)) + \{(g_{Z}(A=1,W)- g_{Z}(A=0,W)) - \Psi_{FS} \}.
\end{equation}


We first solve $D_{SDE}$ to obtain the EE estimate of $\Psi_{SDE}$. We calculate the first component of $D_{SDE}$ as follows. 
Let $g_M=P(M=m | Z,A,W), g_A=P(A=a | W)$, and $g_{A2}=P(A=a | W,Z,M)$. Recall that $\hat{g}_{M|0,W}$ is treated as known, estimated from the observed data, marginalizing out $Z: \sum\limits^1\limits_{z=0} P(M=m|Z=z,A=0,W)P(Z=z|A=0,W)$ \citep{vanderweele2016mediation}. 
  $g_{M,n}$ can be estimated by predicted probabilities from a logistic regression model of $M$ on $Z$, $A$, and $W$. 
  $g_{A2}$ can be written $\frac{P(A=a | W)P(Z | a, W)P(M|Z,a,W)}{P(M,Z | W)} = \frac{g_{A|W}g_{Z|A,W}g_{M|Z,A,W}}{P(M,Z | W)} $, where $g_{A,n}$ and $g_{M,n}$ can be estimated as described above, $g_{Z,n}$ can be estimated from a logistic regression model of $Z$ on $A$ and $W$, and where an estimate of $P(Z,M| W)$ is obtained by marginalizing out $A: \bigg( \sum\limits^1\limits_{a=0} P(M=m|Z,A=a,W)P(A=a | W) \bigg)\bigg(\sum\limits^1\limits_{a=0} P(Z=z|A=a,W)P(A=a | W)\bigg),$ which can be rewritten in terms of the above estimators $ \sum\limits^1\limits_{a=0} g_{M,n}g_{A,n}\sum\limits^1\limits_{a=0} g_{Z,n}g_{A,n}$
  . The other components can be calculated as described in the main text.
 
The second and third components of $D_{SDE}$ and the components of $D_{FS}$ are calculated as described in the main text.  The associated variance can be estimated as the sample variance of the EIC, $D_{CSDE}(P)$, which is given in Equation \ref{eqEEeic}. 

\subsection{Compatible Targeted Minimum Loss-Based Estimator}
Recall $\bar{Q}_Y=E(Y|W,Z,M)$, 
 $g_M=P(M=m | Z,A,W), g_A=P(A=a | W)$, and $g_{A2}=P(A=a | W,Z,M)$. Again, $\hat{g}_{M|0,W}$ is treated as known, estimated from the observed data, marginalizing out $Z: \sum\limits^1\limits_{z=0} P(M=m|Z=z,A=0,W)P(Z=z|A=0,W)$ \citep{vanderweele2016mediation}. Consider submodel $\{\bar{Q}_{Y,n}(M,Z,W)(\epsilon):\epsilon\}$ defined as:
$logit(\bar{Q}_{Y,n} (\epsilon)(M,Z,W)) = logit(\bar{Q}_{Y,n}(M,Z,W) ) + \epsilon C_Y$, where $C_Y = \bigg(\frac{g_{1|W,Z,M}}{g_{1|W}} - \frac{g_{0|W,Z,M}}{g_{0|W}}\bigg) \frac{\hat{g}_{M|A=0,W} }{g_{M|Z,A,W}}$.

The components of $C_Y$ can be calculated as described in the above subsections and in the main text. The update step for $\bar{Q}_Y$ and the remaining steps for the TMLE estimator are completed as in the main text. 

The TMLE solves the efficient influence curve (EIC) for $\Psi_{CSDE}$ (shown in the previous subsection), replacing $g_{Z}$ and $\bar{Q}_{Y}$ with $g_{Z}^*$ and $\bar{Q}^*_{Y}$. The variance of the TMLE estimator of $\Psi_{CSDE}$ is estimated as the sample variance of $D_{CSDE}(P)$. 


\section{R code}
\subsection{Code for ratio of Inverse Probability of Treatment Weighted Estimators}

\begin{lstlisting}
#This estimates the complier stochastic direct effect and its variance. It takes the following arguments:
# a is the instrument, 0/1. It is assumed to be exogenous, but the code can be modified to make it conditionally random.
# z is the exposure influenced by the instrument, 0/1. It is a function of a and w
# m is the mediator, 0/1. It is a function of z, w.
# y is the outcome, 0/1, but the code can be modified for any outcome type. It is a function of z, w, m. 
# w is a matrix of covariates
# svywt is a vector of weights to be applied to the data. 
# zmodel is the parametric model for z.
# mmodel is the parametric model for m.
# ymodel is the parametric model for y. 
# qmodel is the parametric model for q.
# gm is the user-specified stochastic intervention on M, conditional on a=0 and w
# za, za1, and za0 are optional arguments that can be included if the user estimates these as part of the stochastic intervention. Otherwise, they are estimated within the function
# uses the constrained regression function if za, za1, and za0 are null

mediptw<-function(a, z, m, y, w, svywt, zmodel, mmodel, ymodel, qmodel, gm, za=NULL, za1=NULL, za0=NULL){

datw<-w

# estimate p(m | w, z)
mz<-predict(glm(formula=mmodel, family="binomial", data=data.frame(cbind(datw, z=z, m=m))), newdata=data.frame(cbind(datw, z=z)), type="response")
mz0<-predict(glm(formula=mmodel, family="binomial", data=data.frame(cbind(datw, z=z, m=m))), newdata=data.frame(cbind(datw, z=0)), type="response")
mz1<-predict(glm(formula=mmodel, family="binomial", data=data.frame(cbind(datw, z=z, m=m))), newdata=data.frame(cbind(datw, z=1)), type="response")

# estimate p(z | w, a)
if(is.null(za) | is.null(za1) | is.null(za0)){
 zfit<-mle.logreg.constrained(formula(zmodel), data.frame(cbind(datw, a=a, z=z)))

  za0<-predictClogis(cbind(rep(0,nrow(data.frame(datw))), datw), zfit$beta)
  za1<-predictClogis(cbind(rep(1,nrow(data.frame(datw))), datw), zfit$beta)
  za<-predictClogis(data.frame(cbind(a=a, datw)), zfit$beta)
}
else {
  za<-za
  za1<-za1
  za0<-za0
}

pza1<-ifelse(z==1, za1, 1-za1)
pza0<-ifelse(z==1, za0, 1-za0)

# estimate p(a|w,m,z) using previous estimates. Note that p(a|w,m,z) = p(a|w,z) bc of exclusion restriction 
pa1<-(mean(a)*pza1)/(pza1*mean(a) + pza0*mean(1-a))
pa1z0<-(mean(a)*(1-za1))/((1-za1)*mean(a) + (1-za0)*mean(1-a))
pa1z1<-(mean(a)*za1)/(za1*mean(a) + za0*mean(1-a))

  tmpdat<-data.frame(cbind(datw, a=a))

 #make clever covariate
  psm<-(mz*m) + ((1-mz)*(1-m))
  
  tmpdat$wts<-((m*gm + (1-m)*(1-gm))/psm)* svywt
  #component that can't go into the weights
  tmpdat$cc<- (pa1/mean(a))  - ((1-pa1)/mean(1-a))

  tmpdat$ccz0<- (pa1z0/mean(a))  - ((1-pa1z0)/mean(1-a))
  tmpdat$ccz1<- (pa1z1/mean(a))  - ((1-pa1z1)/mean(1-a))
 
  tmpdat$y<-y
  
  psi1<-sum(tmpdat$y * tmpdat$wts * tmpdat$cc)/sum(svywt) 
  eicpsi1<-(tmpdat$cc*tmpdat$wts * tmpdat$y) - psi1

  #estimate denominator 
  psi2<-sum(z * tmpdat$cc *svywt)/sum(svywt)
  eicpsi2<-(z * tmpdat$cc * svywt) - psi2

  csde<-psi1/psi2
  csdeeic<-(eicpsi1/psi2) - ((psi1*eicpsi2)/(psi2^2))
  varcsde<-var(csdeeic)/nrow(tmpdat)

 return(list("est"=csde, "var"=varcsde))
}
\end{lstlisting}

\subsection{Code for ratio of Estimating Equation Estimators}

\begin{lstlisting}
#This estimates the complier stochastic direct effect and its variance. It takes the following arguments:
# a is the instrument, 0/1. It is assumed to be exogenous, but the code can be modified to make it conditionally random.
# z is the exposure influenced by the instrument, 0/1. It is a function of a and w
# m is the mediator, 0/1. It is a function of z, w.
# y is the outcome, 0/1, but the code can be modified for any outcome type. It is a function of z, w, m. 
# w is a matrix of covariates
# svywt is a vector of weights to be applied to the data. 
# zmodel is the parametric model for z.
# mmodel is the parametric model for m.
# ymodel is the parametric model for y. 
# qmodel is the parametric model for q.
# gm is the user-specified stochastic intervention on M, conditional on a=0 and w
# za, za1, and za0 are optional arguments that can be included if the user estimates these as part of the stochastic intervention. Otherwise, they are estimated within the function
# uses the constrained regression function if za, za1, and za0 are null

medee<-function(a, z, m, y, w, svywt, zmodel, mmodel, ymodel, qmodel, gm, za=NULL, za1=NULL, za0=NULL){

datw<-w

# estimate p(m | w, z)
mz<-predict(glm(formula=mmodel, family="binomial", data=data.frame(cbind(datw, z=z, m=m))), newdata=data.frame(cbind(datw, z=z)), type="response")
mz0<-predict(glm(formula=mmodel, family="binomial", data=data.frame(cbind(datw, z=z, m=m))), newdata=data.frame(cbind(datw, z=0)), type="response")
mz1<-predict(glm(formula=mmodel, family="binomial", data=data.frame(cbind(datw, z=z, m=m))), newdata=data.frame(cbind(datw, z=1)), type="response")

# estimate p(z | w, a)
if(is.null(za) | is.null(za1) | is.null(za0)){
 zfit<-mle.logreg.constrained(formula(zmodel), data.frame(cbind(datw, a=a, z=z)))

  za0<-predictClogis(cbind(rep(0,nrow(data.frame(datw))), datw), zfit$beta)
  za1<-predictClogis(cbind(rep(1,nrow(data.frame(datw))), datw), zfit$beta)
  za<-predictClogis(data.frame(cbind(a=a, datw)), zfit$beta)
}
else {
  za<-za
  za1<-za1
  za0<-za0
}

pza1<-ifelse(z==1, za1, 1-za1)
pza0<-ifelse(z==1, za0, 1-za0)

# estimate p(a|w,m,z) using previous estimates. Note that p(a|w,m,z) = p(a|w,z) bc of exclusion restriction 
pa1<-(mean(a)*pza1)/(pza1*mean(a) + pza0*mean(1-a))
pa1z0<-(mean(a)*(1-za1))/((1-za1)*mean(a) + (1-za0)*mean(1-a))
pa1z1<-(mean(a)*za1)/(za1*mean(a) + za0*mean(1-a))

  tmpdat<-data.frame(cbind(datw, a=a))

 #get initial Y fit
  yfit<-glm(formula=ymodel, family="binomial", data=data.frame(cbind(datw, z=z, m=m, y=y)))
  tmpdat$qyinit<-cbind(predict(yfit, newdata=data.frame(cbind(datw, z=z, m=m)), type="response"), 
    predict(yfit, newdata=data.frame(cbind(datw, z=z, m=1)), type="response"), 
    predict(yfit, newdata=data.frame(cbind(datw, z=z, m=0)), type="response") )
  tmpdat$qyinitz0<-cbind(predict(yfit, newdata=data.frame(cbind(datw, z=0, m=1)), type="response"), predict(yfit, newdata=data.frame(cbind(datw, z=0, m=0)), type="response") )
  tmpdat$qyinitz1<-cbind(predict(yfit, newdata=data.frame(cbind(datw, z=1, m=1)), type="response"), predict(yfit, newdata=data.frame(cbind(datw, z=1, m=0)), type="response") )

#make clever covariate
  psm<-(mz*m) + ((1-mz)*(1-m))
  
  tmpdat$wts<-((m*gm + (1-m)*(1-gm))/psm)* svywt
  #component that can't go into the weights
  tmpdat$cc<- (pa1/mean(a))  - ((1-pa1)/mean(1-a))

  tmpdat$ccz0<- (pa1z0/mean(a))  - ((1-pa1z0)/mean(1-a))
  tmpdat$ccz1<- (pa1z1/mean(a))  - ((1-pa1z1)/mean(1-a))
 
  tmpdat$y<-y
  
  epsilon<-coef(glm(y ~  -1 + offset(qlogis(qyinit[,1])) +cc ,weights=wts, family="quasibinomial", data=tmpdat))  #
 
  eic1<-(tmpdat$cc)*tmpdat$wts * (tmpdat$y - tmpdat$qyinit[,1])

  #integrate out M to get qm
  tmpdat$qm<-tmpdat$qyinit[,2]*gm  + tmpdat$qyinit[,3]*(1-gm)
  tmpdat$qmz0<-tmpdat$qyinitz0[,1]*gm  + tmpdat$qyinitz0[,2]*(1-gm)
  tmpdat$qmz1<-tmpdat$qyinitz1[,1]*gm  + tmpdat$qyinitz1[,2]*(1-gm)

  #initial fit gz
  gz<-cbind(za, za0, za1)
    
  #make components for second targeting step
  tmpdat$difqmza<-tmpdat$qmz1 - tmpdat$qmz0
  
  tmpdat$ga<-ifelse(a==1, mean(a), mean(1-a))
  tmpdat$a<-a
  tmpdat$nota<-1-a

  #integrate out z to get qz
  qz<-cbind((tmpdat$qmz1*gz[,2]) + (tmpdat$qmz0*(1-gz[,2])), (tmpdat$qmz1*gz[,3]) + (tmpdat$qmz0*(1-gz[,3])))
  
  #estimate numerator
  eic2<-((2*a-1)/tmpdat$ga)*svywt*(tmpdat$qmz1 - tmpdat$qmz0) *(z-gz[,1])
  #eic2<-(tmpdat$a*svywt*(tmpdat$qmz1 - tmpdat$qmz0) + tmpdat$nota*svywt*(tmpdat$qmz1 - tmpdat$qmz0))*(z-gz[,1])

  #estimate denominator 
  eic3<-((qz[,2] - qz[,1])*svywt)
  psi1<-mean(eic1 + eic2+ eic3)
  eicdp1<-eic1 + eic2 + eic3 -psi1
  
  eic1dp2<-((2*a-1)/tmpdat$ga)*svywt*(z - gz[,1])
  eic2dp2<-((gz[,3] - gz[,2])*svywt) 
  psi2<-mean(eic1dp2 + eic2dp2)
  eicdp2<-eic1dp2 + eic2dp2 - psi2

  
  csde<-psi1/psi2
  csdeeic<-(eicdp1/psi2) - ((psi1*eicdp2)/(psi2^2))
  varcsde<-var(csdeeic)/nrow(tmpdat)

 return(list("est"=csde, "var"=varcsde))
}
\end{lstlisting}

\subsection{Code for ratio of Targeted Minimum Loss-based  Estimators (Efficient TMLE)}

\begin{lstlisting}
#This estimates the complier stochastic direct effect and its variance. It takes the following arguments:
# a is the instrument, 0/1. It is assumed to be exogenous, but the code can be modified to make it conditionally random.
# z is the exposure influenced by the instrument, 0/1. It is a function of a and w
# m is the mediator, 0/1. It is a function of z, w.
# y is the outcome, 0/1, but the code can be modified for any outcome type. It is a function of z, w, m. 
# w is a matrix of covariates
# svywt is a vector of weights to be applied to the data. 
# zmodel is the parametric model for z.
# mmodel is the parametric model for m.
# ymodel is the parametric model for y. 
# qmodel is the parametric model for q.
# gm is the user-specified stochastic intervention on M, conditional on a=0 and w
# za, za1, and za0 are optional arguments that can be included if the user estimates these as part of the stochastic intervention. Otherwise, they are estimated within the function
# uses the constrained regression function if za, za1, and za0 are null

medtmle<-function(a, z, m, y, w, svywt, zmodel, mmodel, ymodel, qmodel, gm, za=NULL, za1=NULL, za0=NULL){

datw<-w

# estimate p(m | w, z)
mz<-predict(glm(formula=mmodel, family="binomial", data=data.frame(cbind(datw, z=z, m=m))), newdata=data.frame(cbind(datw, z=z)), type="response")
mz0<-predict(glm(formula=mmodel, family="binomial", data=data.frame(cbind(datw, z=z, m=m))), newdata=data.frame(cbind(datw, z=0)), type="response")
mz1<-predict(glm(formula=mmodel, family="binomial", data=data.frame(cbind(datw, z=z, m=m))), newdata=data.frame(cbind(datw, z=1)), type="response")

# estimate p(z | w, a)
if(is.null(za) | is.null(za1) | is.null(za0)){
 zfit<-mle.logreg.constrained(formula(zmodel), data.frame(cbind(datw, a=a, z=z)))

  za0<-predictClogis(cbind(rep(0,nrow(data.frame(datw))), datw), zfit$beta)
  za1<-predictClogis(cbind(rep(1,nrow(data.frame(datw))), datw), zfit$beta)
  za<-predictClogis(data.frame(cbind(a=a, datw)), zfit$beta)
}
else {
  za<-za
  za1<-za1
  za0<-za0
}

pza1<-ifelse(z==1, za1, 1-za1)
pza0<-ifelse(z==1, za0, 1-za0)

# estimate p(a|w,m,z) using previous estimates. Note that p(a|w,m,z) = p(a|w,z) bc of exclusion restriction 
pa1<-(mean(a)*pza1)/(pza1*mean(a) + pza0*mean(1-a))
pa1z0<-(mean(a)*(1-za1))/((1-za1)*mean(a) + (1-za0)*mean(1-a))
pa1z1<-(mean(a)*za1)/(za1*mean(a) + za0*mean(1-a))

  tmpdat<-data.frame(cbind(datw, a=a))

 #get initial Y fit
  yfit<-glm(formula=ymodel, family="binomial", data=data.frame(cbind(datw, z=z, m=m, y=y)))
  tmpdat$qyinit<-cbind(predict(yfit, newdata=data.frame(cbind(datw, z=z, m=m)), type="response"), 
    predict(yfit, newdata=data.frame(cbind(datw, z=z, m=1)), type="response"), 
    predict(yfit, newdata=data.frame(cbind(datw, z=z, m=0)), type="response") )
  tmpdat$qyinitz0<-cbind(predict(yfit, newdata=data.frame(cbind(datw, z=0, m=1)), type="response"), predict(yfit, newdata=data.frame(cbind(datw, z=0, m=0)), type="response") )
  tmpdat$qyinitz1<-cbind(predict(yfit, newdata=data.frame(cbind(datw, z=1, m=1)), type="response"), predict(yfit, newdata=data.frame(cbind(datw, z=1, m=0)), type="response") )

#make clever covariate
  psm<-(mz*m) + ((1-mz)*(1-m))
  
  tmpdat$wts<-((m*gm + (1-m)*(1-gm))/psm)* svywt
  #component that can't go into the weights
  tmpdat$cc<- (pa1/mean(a))  - ((1-pa1)/mean(1-a))

  tmpdat$ccz0<- (pa1z0/mean(a))  - ((1-pa1z0)/mean(1-a))
  tmpdat$ccz1<- (pa1z1/mean(a))  - ((1-pa1z1)/mean(1-a))
 
  tmpdat$y<-y
  
  epsilon<-coef(glm(y ~  -1 + offset(qlogis(qyinit[,1])) +cc ,weights=wts, family="quasibinomial", data=tmpdat))  #
 epsilon<-ifelse(is.na(epsilon), 0, epsilon)

 #update Qy
  tmpdat$qyup<-plogis(qlogis(tmpdat$qyinit) + epsilon*(tmpdat$cc))
  tmpdat$qyupz0m0<-plogis(qlogis(tmpdat$qyinitz0[,2]) + epsilon * tmpdat$ccz0)
  tmpdat$qyupz0m1<-plogis(qlogis(tmpdat$qyinitz0[,1]) + epsilon * tmpdat$ccz0)
  tmpdat$qyupz1m0<-plogis(qlogis(tmpdat$qyinitz1[,2]) + epsilon * tmpdat$ccz1)
  tmpdat$qyupz1m1<-plogis(qlogis(tmpdat$qyinitz1[,1]) + epsilon * tmpdat$ccz1)
  
  eic1<-(tmpdat$cc)*tmpdat$wts * (tmpdat$y - tmpdat$qyup[,1])

  #integrate out M to get qm
  tmpdat$qm<-tmpdat$qyup[,2]*gm  + tmpdat$qyup[,3]*(1-gm)
  tmpdat$qmz0<-tmpdat$qyupz0m1*gm  + tmpdat$qyupz0m0*(1-gm)
  tmpdat$qmz1<-tmpdat$qyupz1m1*gm  + tmpdat$qyupz1m0*(1-gm)

  #initial fit gz
  gz<-cbind(za, za0, za1)
    
  #make components for second targeting step
  tmpdat$difqmza<-tmpdat$qmz1 - tmpdat$qmz0
  
  tmpdat$ga<-ifelse(a==1, mean(a), mean(1-a))
  tmpdat$a<-a
  tmpdat$nota<-1-a

  fitcz<- glm(z ~ -1 + a:difqmza + nota:difqmza, weights=svywt*(1/tmpdat$ga), family="quasibinomial", data=tmpdat, offset=qlogis(gz[,1]))

  epsiloncz<-coef(fitcz)
  
  #update gz
  gzup<-cbind(plogis(qlogis(gz[,1]) + I(tmpdat$a==0)*epsiloncz[2]*tmpdat$difqmza + I(tmpdat$a==1)*epsiloncz[1]*tmpdat$difqmza),
    plogis(qlogis(gz[,2]) +  epsiloncz[2]*tmpdat$difqmza),
    plogis(qlogis(gz[,3]) + epsiloncz[1]*tmpdat$difqmza)
    )

  #integrate out z to get qz
  qz<-cbind((tmpdat$qmz1*gzup[,2]) + (tmpdat$qmz0*(1-gzup[,2])), (tmpdat$qmz1*gzup[,3]) + (tmpdat$qmz0*(1-gzup[,3])))
  
  #estimate numerator
  psi1<-sum((qz[,2]-qz[,1])*svywt)/sum(svywt)
 
  #eic2<-(tmpdat$a*svywt*(tmpdat$qmz1 - tmpdat$qmz0) + tmpdat$nota*svywt*(tmpdat$qmz1 - tmpdat$qmz0))*(z-gzup[,1])
  eic2<-((2*tmpdat$a-1)/tmpdat$ga)*svywt*(tmpdat$qmz1 - tmpdat$qmz0) *(z-gzup[,1])
  #target gz for denominator
  fitczd<- glm(z ~ -1 + a + nota, weights=svywt*(1/tmpdat$ga), family="quasibinomial", data=tmpdat, offset=qlogis(gz[,1]))
  epsilonczd<-coef(fitczd)
  gzupd<-cbind(plogis(qlogis(gz[,1]) + I(tmpdat$a==0)*epsilonczd[2] + I(tmpdat$a==1)*epsilonczd[1]),
    plogis(qlogis(gz[,2]) +  epsilonczd[2]),
    plogis(qlogis(gz[,3]) + epsilonczd[1])
    )
  #estimate denominator 
  psi2<-sum((gzupd[,3]-gzupd[,2])*svywt)/sum(svywt)

 
  eic3<-((qz[,2] - qz[,1])*svywt) - psi1
  eicdp1<-eic1 + eic2+ eic3

  eic1dp2<-((2*a-1)/tmpdat$ga)*svywt*(z - gz[,1])
  eic2dp2<-((gzupd[,3] - gzupd[,2])*svywt) - psi2
  eicdp2<-eic1dp2 + eic2dp2
  
  csde<-psi1/psi2
  csdeeic<-(eicdp1/psi2) - ((psi1*eicdp2)/(psi2^2))
  varcsde<-var(csdeeic)/nrow(tmpdat)

 return(list("est"=csde, "var"=varcsde))
}
\end{lstlisting}

\subsection{Code for Targeted Minimum Loss-based  Estimator that estimates ratio directly (Compatible TMLE)}

\begin{lstlisting}
#This estimates the complier stochastic direct effect and its variance. It takes the following arguments:
# a is the instrument, 0/1. It is assumed to be exogenous, but the code can be modified to make it conditionally random.
# z is the exposure influenced by the instrument, 0/1. It is a function of a and w
# m is the mediator, 0/1. It is a function of z, w.
# y is the outcome, 0/1, but the code can be modified for any outcome type. It is a function of z, w, m. 
# w is a matrix of covariates
# svywt is a vector of weights to be applied to the data. 
# zmodel is the parametric model for z.
# mmodel is the parametric model for m.
# ymodel is the parametric model for y. 
# qmodel is the parametric model for q.
# gm is the user-specified stochastic intervention on M, conditional on a=0 and w
# za, za1, and za0 are optional arguments that can be included if the user estimates these as part of the stochastic intervention. Otherwise, they are estimated within the function
# uses the constrained regression function if za, za1, and za0 are null

medtmle<-function(a, z, m, y, w, svywt, zmodel, mmodel, ymodel, qmodel, gm, za=NULL, za1=NULL, za0=NULL){

datw<-w

# estimate p(m | w, z)
mz<-predict(glm(formula=mmodel, family="binomial", data=data.frame(cbind(datw, z=z, m=m))), newdata=data.frame(cbind(datw, z=z)), type="response")
mz0<-predict(glm(formula=mmodel, family="binomial", data=data.frame(cbind(datw, z=z, m=m))), newdata=data.frame(cbind(datw, z=0)), type="response")
mz1<-predict(glm(formula=mmodel, family="binomial", data=data.frame(cbind(datw, z=z, m=m))), newdata=data.frame(cbind(datw, z=1)), type="response")

# estimate p(z | w, a)
if(is.null(za) | is.null(za1) | is.null(za0)){
 zfit<-mle.logreg.constrained(formula(zmodel), data.frame(cbind(datw, a=a, z=z)))

  za0<-predictClogis(cbind(rep(0,nrow(data.frame(datw))), datw), zfit$beta)
  za1<-predictClogis(cbind(rep(1,nrow(data.frame(datw))), datw), zfit$beta)
  za<-predictClogis(data.frame(cbind(a=a, datw)), zfit$beta)
}
else {
  za<-za
  za1<-za1
  za0<-za0
}

pza1<-ifelse(z==1, za1, 1-za1)
pza0<-ifelse(z==1, za0, 1-za0)

# estimate p(a|w,m,z) using previous estimates. Note that p(a|w,m,z) = p(a|w,z) bc of exclusion restriction 
pa1<-(mean(a)*pza1)/(pza1*mean(a) + pza0*mean(1-a))
pa1z0<-(mean(a)*(1-za1))/((1-za1)*mean(a) + (1-za0)*mean(1-a))
pa1z1<-(mean(a)*za1)/(za1*mean(a) + za0*mean(1-a))

  tmpdat<-data.frame(cbind(datw, a=a))

 #get initial Y fit
  yfit<-glm(formula=ymodel, family="binomial", data=data.frame(cbind(datw, z=z, m=m, y=y)))
  tmpdat$qyinit<-cbind(predict(yfit, newdata=data.frame(cbind(datw, z=z, m=m)), type="response"), 
    predict(yfit, newdata=data.frame(cbind(datw, z=z, m=1)), type="response"), 
    predict(yfit, newdata=data.frame(cbind(datw, z=z, m=0)), type="response") )
  tmpdat$qyinitz0<-cbind(predict(yfit, newdata=data.frame(cbind(datw, z=0, m=1)), type="response"), predict(yfit, newdata=data.frame(cbind(datw, z=0, m=0)), type="response") )
  tmpdat$qyinitz1<-cbind(predict(yfit, newdata=data.frame(cbind(datw, z=1, m=1)), type="response"), predict(yfit, newdata=data.frame(cbind(datw, z=1, m=0)), type="response") )

#make clever covariate
  psm<-(mz*m) + ((1-mz)*(1-m))
  
  tmpdat$wts<-((m*gm + (1-m)*(1-gm))/psm)* svywt
  #component that can't go into the weights
  tmpdat$cc<- (pa1/mean(a))  - ((1-pa1)/mean(1-a))

  tmpdat$ccz0<- (pa1z0/mean(a))  - ((1-pa1z0)/mean(1-a))
  tmpdat$ccz1<- (pa1z1/mean(a))  - ((1-pa1z1)/mean(1-a))
 
  tmpdat$y<-y
  
  epsilon<-coef(glm(y ~  -1 + offset(qlogis(qyinit[,1])) +cc ,weights=wts, family="quasibinomial", data=tmpdat))  #
 
 #update Qy
  tmpdat$qyup<-plogis(qlogis(tmpdat$qyinit) + epsilon*(tmpdat$cc))
  tmpdat$qyupz0m0<-plogis(qlogis(tmpdat$qyinitz0[,2]) + epsilon * tmpdat$ccz0)
  tmpdat$qyupz0m1<-plogis(qlogis(tmpdat$qyinitz0[,1]) + epsilon * tmpdat$ccz0)
  tmpdat$qyupz1m0<-plogis(qlogis(tmpdat$qyinitz1[,2]) + epsilon * tmpdat$ccz1)
  tmpdat$qyupz1m1<-plogis(qlogis(tmpdat$qyinitz1[,1]) + epsilon * tmpdat$ccz1)
  
  eic1<-(tmpdat$cc)*tmpdat$wts * (tmpdat$y - tmpdat$qyup[,1])

  #integrate out M to get qm
  tmpdat$qm<-tmpdat$qyup[,2]*gm  + tmpdat$qyup[,3]*(1-gm)
  tmpdat$qmz0<-tmpdat$qyupz0m1*gm  + tmpdat$qyupz0m0*(1-gm)
  tmpdat$qmz1<-tmpdat$qyupz1m1*gm  + tmpdat$qyupz1m0*(1-gm)

  #initial fit gz
  gz<-cbind(za, za0, za1)
    
  #make components for second targeting step
  tmpdat$difqmza<-tmpdat$qmz1 - tmpdat$qmz0
  
  tmpdat$ga<-ifelse(a==1, mean(a), mean(1-a))
  tmpdat$a<-a
  tmpdat$nota<-1-a

  fitcz<- glm(z ~ -1 + a + nota + a:difqmza + nota:difqmza, weights=svywt*(1/tmpdat$ga), family="quasibinomial", data=tmpdat, offset=qlogis(gz[,1]))

  epsiloncz<-coef(fitcz)
  
  #update gz
  gzup<-cbind(plogis(qlogis(gz[,1]) + I(tmpdat$a==0)*epsiloncz[2] + I(tmpdat$a==0)*epsiloncz[4]*tmpdat$difqmza + I(tmpdat$a==1)*epsiloncz[1] + I(tmpdat$a==1)*epsiloncz[3]*tmpdat$difqmza),
    plogis(qlogis(gz[,2]) + epsiloncz[2] + epsiloncz[4]*tmpdat$difqmza),
    plogis(qlogis(gz[,3]) + epsiloncz[1] + epsiloncz[3]*tmpdat$difqmza)
    )

  #integrate out z to get qz
  qz<-cbind((tmpdat$qmz1*gzup[,2]) + (tmpdat$qmz0*(1-gzup[,2])), (tmpdat$qmz1*gzup[,3]) + (tmpdat$qmz0*(1-gzup[,3])))
  
  #estimate numerator
  psi1<-sum((qz[,2]-qz[,1])*svywt)/sum(svywt)
 
  eic2<-((2*a-1)/tmpdat$ga)*svywt*(tmpdat$qmz1 - tmpdat$qmz0) *(z-gzup[,1])
  #estimate denominator 
  psi2<-sum((gzup[,3]-gzup[,2])*svywt)/sum(svywt)
 
  eic3<-((qz[,2] - qz[,1])*svywt) - psi1
  eicdp1<-eic1 + eic2+ eic3

  eic1dp2<-((2*a-1)/tmpdat$ga)*svywt*(z - gz[,1])
  eic2dp2<-((gzup[,3] - gzup[,2])*svywt) - psi2
  eicdp2<-eic1dp2 + eic2dp2
  
  csde<-psi1/psi2
  csdeeic<-(eicdp1/psi2) - ((psi1*eicdp2)/(psi2^2))
  varcsde<-var(csdeeic)/nrow(tmpdat)

 return(list("est"=csde, "var"=varcsde))
}
\end{lstlisting}

\section{Figures for stochastic direct and indirect effects}
\begin{figure}[H]
\centerline{\includegraphics[width=6in]{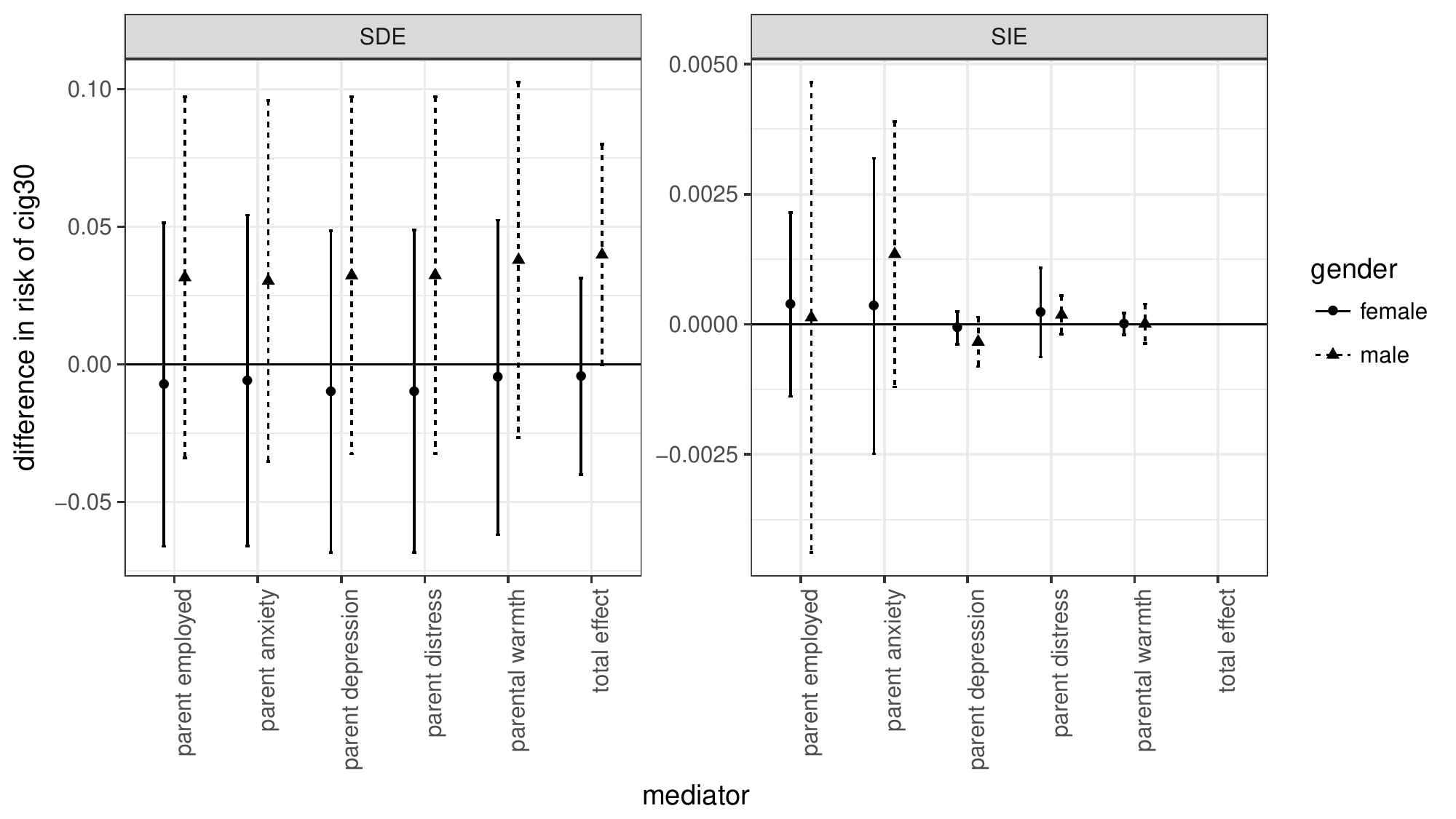}}
\caption{Data-dependent stochastic direct and indirect effect estimates and 95\% confidence intervals on past-month cigarette use by mediator. Data from the Moving to Opportunity experiment, interim follow up. 
}\label{cigplotsdesie}

\end{figure}

\begin{figure}[H]
\centerline{\includegraphics[width=6in]{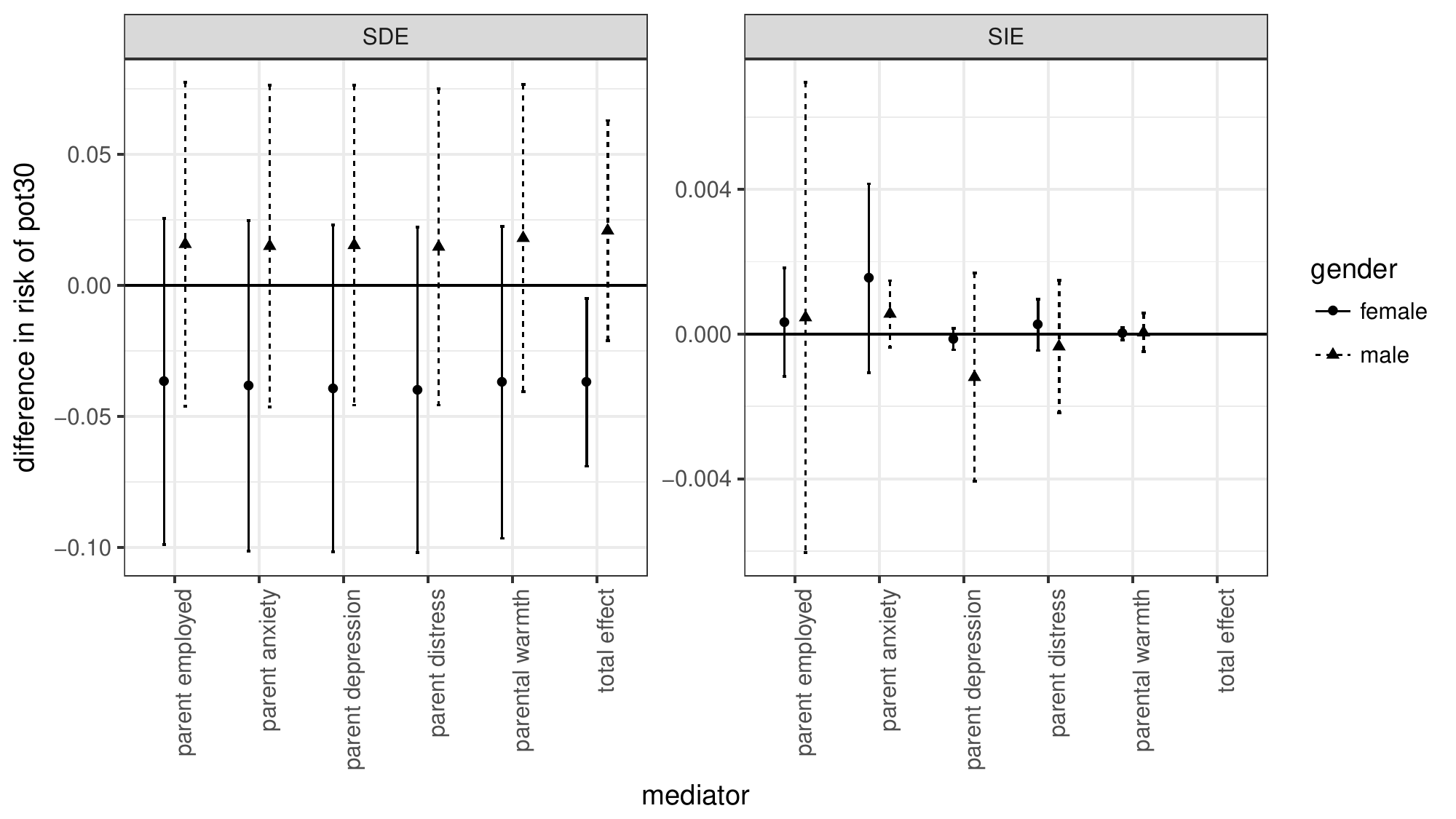}}
\caption{Data-dependent stochastic direct and indirect effect estimates and 95\% confidence intervals on past-month marijuana use by mediator. Data from the Moving to Opportunity experiment, interim follow up.
}\label{potplotsdesie}
\end{figure}

\begin{figure}[H]
\centerline{\includegraphics[width=6in]{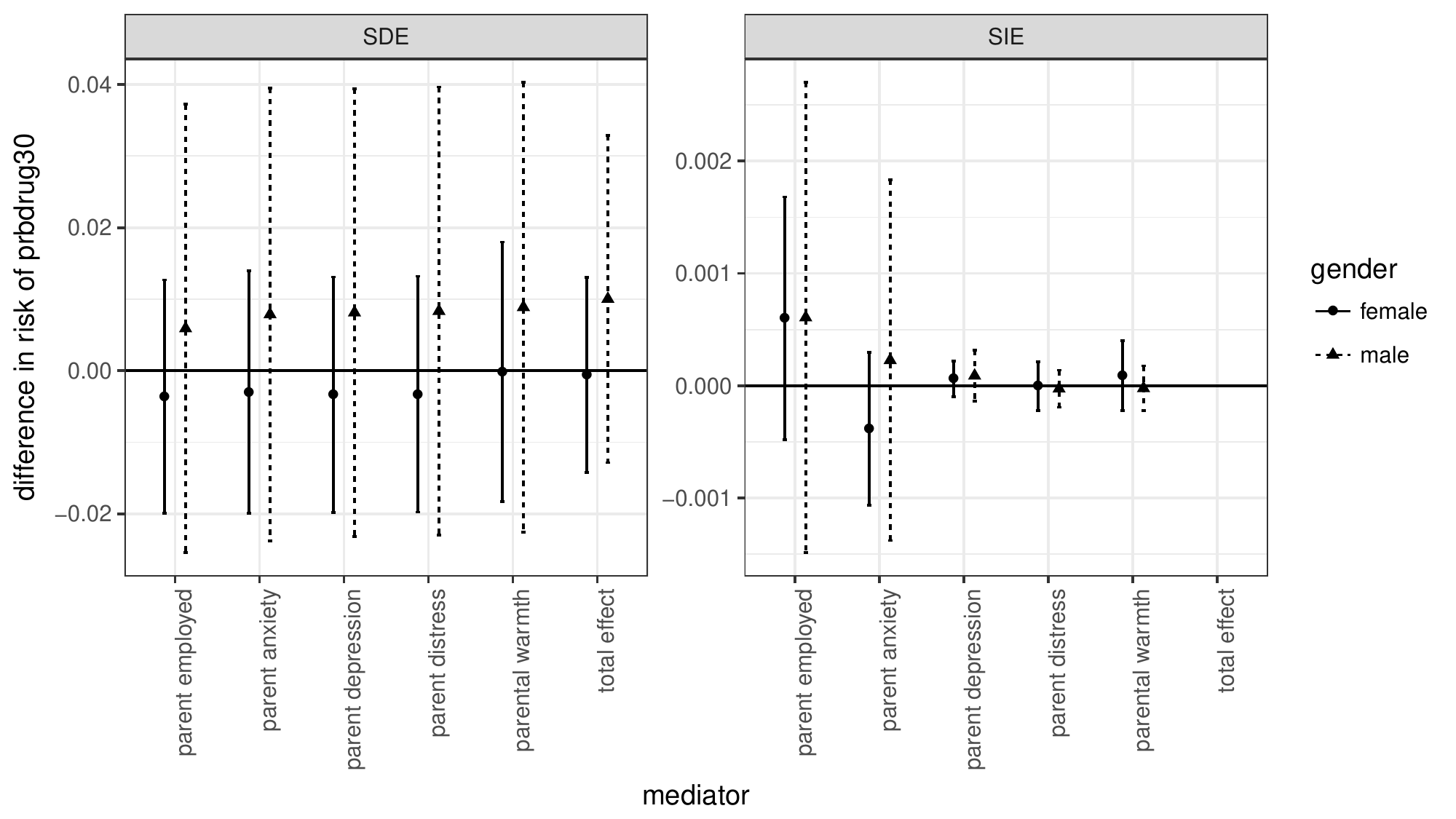}}
\caption{Data-dependent stochastic direct and indirect effect estimates and 95\% confidence intervals on past-month problematic drug use by mediator. Data from the Moving to Opportunity experiment, interim follow up. 
}\label{prbdrugplotsdesie}
\end{figure}

\end{document}